\documentclass[%
 reprint,
superscriptaddress,
 amsmath,amssymb,
 aps,
]{revtex4-2}

\usepackage{graphicx}
\usepackage{dcolumn}
\usepackage{bm}
\usepackage{bbm}
\usepackage{esint}
\usepackage{physics}
\usepackage{hyperref}

\usepackage{color}
\newcommand{\edited}[1]{\textcolor{blue}{#1}}
\newcommand{\neededit}[1]{\textcolor{red}{#1}}

\begin{document}

\preprint{APS/123-QED}

\title{Time-reversal-symmetry bounds on electromagnetic fields}

\author{Wenchao~Ma}
\affiliation{Department of Chemistry, Massachusetts Institute of Technology, Cambridge, MA 02139, USA}
\author{Rapha{\"e}l~Pestourie}
\affiliation{School of Computational Science and Engineering, Georgia Institute of Technology, Atlanta, GA 30332, USA}
\author{Steven G. Johnson}
\affiliation{Department of Mathematics, Massachusetts Institute of Technology, Cambridge, MA 02139, USA}
\email{stevenj@math.mit.edu}

\date{\today}

\begin{abstract}
For linear electromagnetic systems possessing time-reversal symmetry, we present an approach to bound ratios of internal fields excited from different ports, using only the scattering matrix ($S$~matrix), improving upon previous related bounds by Sounas and Al{\`{u}} [Phys. Rev. Lett. {\bf{118}}, 154302 (2017)].
By reciprocity, emitted-wave amplitudes from internal dipole sources are bounded in a similar way. When applied to coupled-resonant systems, our method constrains ratios of resonant coupling/decay coefficients. We also obtain a relation for the relative phase of fields excited from the two ports and the ratio of field intensities in a two-port system. In addition, although lossy systems do not have time-reversal symmetry, we can still approximately bound loss-induced non-unitarity of the $S$~matrix using only the lossless $S$~matrix.
We show numerical validations of the near-tightness of our bounds in various scattering systems.
\end{abstract}

\maketitle

\section{Introduction}
In this paper, we introduce a method to bound the internal fields of linear scattering systems with time-reversal symmetry, given only the $S$~matrix relating incoming and outgoing wave modes (Fig.~\ref{fig1}), often yielding significantly tighter bounds than a precursor approach~\cite{Sounas2017} while also extending to bounds on more general quantities.   Our central results are upper and lower bounds~(\ref{eq:Rayleigh-bounds}) on ratios of field amplitudes excited by different ports (Sec.~\ref{sec:trs}), or equivalently (by reciprocity), emitted-wave amplitudes from a point dipole source into different ports (Sec.~\ref{subsec:emit}). Compared with Ref.~\citenum{Sounas2017}, our work incorporates more information from the $S$~matrix in order to tighten the bounds, and we also extend them to more general quadratic ratios of linear functions of the fields.  The near-tightness of our bounds is numerically validated in a variety of 2d and 3d electromagnetic scattering systems (Sec.~\ref{sec:verification} and Fig.~\ref{fig2}), including a non-reciprocal example [Fig.~\ref{fig2}(e)(f)].  We also obtain bounds on ratios of resonant decay/coupling coefficients in coupled-resonant systems (Sec.~\ref{sec:tcmt}), with previous two-port results~\cite{Wang2013} as a special case. We also uncover equalities~(\ref{eq:2port-trs-circle}), (\ref{eq:2port-trs-recipr-circle}), (\ref{eq:tcmt-2port-trs-bounds-k}), and (\ref{eq:tcmt-2port-trs-bounds-d}) relating amplitude ratios and phase differences in two-port systems (Sec.~\ref{sec:2port})~\cite{Wang2013}.  Finally, although time-reversal symmetry typically prohibits/neglects loss (unless it is paired with gain in a non-reciprocal system~\cite{Buddhiraju2020}), we show in Sec.~\ref{sec:absorption} that we can also bound the loss-induced non-unitarity of the $S$~matrix to first order in the loss tangent, given only the \emph{lossless} $S$~matrix.

There has been great recent interest and progress on bounds on electromagnetic responses~\cite{Miller2023,Chao2022}, driven in part by the ability of large-scale optimization (``inverse design'')~\cite{Molesky2018,Christiansen2021} to explore a vast space of parameters for the design of scattering systems.  Bounds help inform the design process by delineating the attainable performance and indicating how much additional performance could potentially be gained over existing designs.   Such upper bounds come in many different forms, depending upon what quantity is being bounded and what information is supplied.  For example, given only the materials and the design volume, one can bound absorption/scattering cross sections~\cite{Miller2016,Gustafsson2020} or emitted powers~\cite{Miller2016,Chao2022nov}.   The bounds we consider in this paper are of a different flavor: given information about the scattered waves, but \emph{no} information on the materials other than their time-reversal symmetry and linearity, what constraints are there on the internal fields and related quantities?   We were inspired by pioneering work on such questions by Sounas and Alu~\cite{Sounas2017}. Their relevant results show that, given a pair of ports in a linear system with time-reversal symmetry and energy conservation, inequalities can be derived for internal fields and the $S$~parameter connecting the two ports. Our work differs in several aspects. First, our method exploits the entire $S$~matrix instead of a single $S$~parameter, so that our bounds are usually tighter. Second, our method can generate various inequalities and equalities for different quantities involving fields excited from multiple ports. Third, only time-reversal symmetry is needed to formulate our theory, while energy conservation (lack of dissipation/gain) and Lorentz reciprocity, if satisified, can be readily incorporated.

As a special case, scattering in a coupled-resonant system is also subject to our bounds, if the system has time-reversal symmetry and is linear.  For sufficiently long-lifetime resonances, coupled-resonant systems can often be modeled by temporal coupled-mode theory (TCMT), in which scattering is split to two pathways: a direct pathway and a resonance-mediated indirect pathway~\cite{Haus1984,Fan2003}. TCMT describes such resonant systems generically using a minimal set of ordinary differential equations whose terms are constrained by general principles such as time-reversal symmetry or reciprocity. The resonances are excited from input ports and decay to output ports, which are quantified by resonant coupling/decay coefficients in TCMT.
Time-reversal symmetry relates these coefficients to the direct-pathway $S$~matrix~\cite{Fan2003}. In this work, expressed in terms of the direct-pathway $S$~matrix, our bounds reveal latent constraints on the resonant coupling and decay coefficients and include some previous results~\cite{Wang2013} as a special case. A different extension to Ref.~\citenum{Wang2013} was recently presented in Ref.~\citenum{Wang2024}.

As the starting point of this work, time-reversal symmetry underlies all of our results. Time reversal is a transformation $t\to-t$ that reverses the arrow of time~\cite{Jackson1998,Sozzi2008,Arntzenius2009,Sigwarth2022,Sakurai2020,Bernabu2015,Lamb1998}. In electromagnetism, this operation reverses the dynamics of fields and sources, and requires material properties to be transformed accordingly (e.g., loss becomes gain) in order for the macroscopic Maxwell equations to still be satisfied. If the time-reversed materials are identical to the original materials, the system has time-reversal symmetry. For time-independent materials in frequency domain, time-reversal symmetry requires permittivity and permeability tensors to be real, and magnetoelectric coupling tensors to be purely imaginary if such couplings are nonzero as in bianisotropic materials~\cite{Silveirinha2019}. Time-reversal symmetry also imposes constraints on the $S$~matrix of a scattering device [Fig.~\ref{fig1}(a)], where $S$ relates amplitudes of input (${\bf a}_{\rm in}$) and output (${\bf a}_{\rm out}$) modes via ${\bf a}_{\rm out}=S{\bf a}_{\rm in}$~\cite{Saleh2019}. If a system has time-reversal symmetry, $S$ satisfies $S^*S=\mathbbm{1}$~\cite{Haus1984}, where $*$ denotes complex conjugation (\emph{not} conjugate transposition) and $\mathbbm{1}$ is the identity matrix. An $S$~matrix explicitly describes externally visible behavior of the system by encapsulating the mapping from input to output waves, but as this work shows, it also implicitly constrains \emph{internal} electromagnetic fields excited from different ports~\cite{Sounas2017}, as well as resonant coupling and decay coefficients in a coupled-resonant system in the context of TCMT~\cite{Fan2003,Suh2004,Wang2018,Zhao2019}. This area of research complements previous studies of fundamental limits on electromagnetic responses~\cite{Chao2022,Miller2023}.

\section{Time-reversal-symmetry bounds in a multi-port structure}\label{sec:trs}

\begin{figure}[ht]
\centering
\includegraphics[width=1\linewidth]{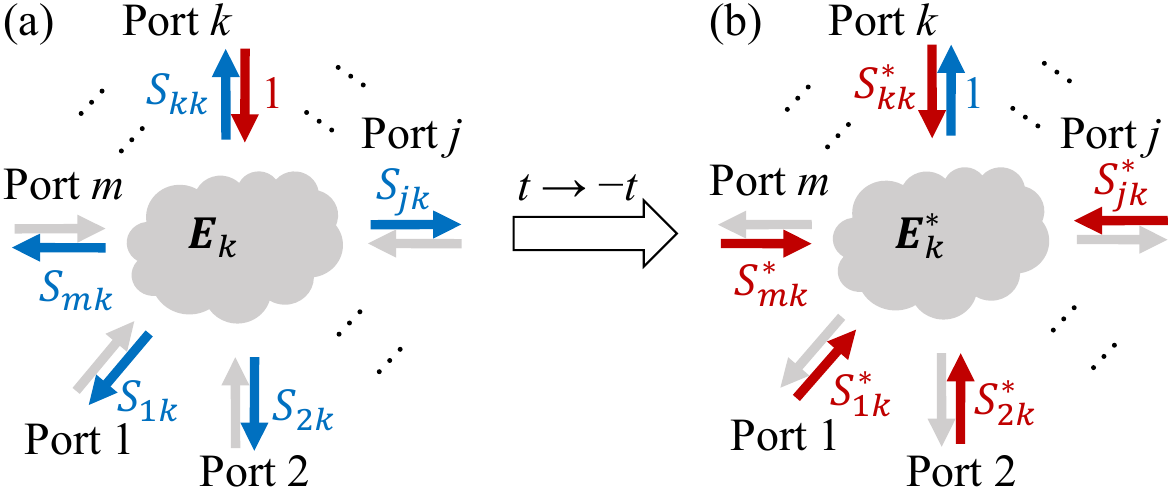}
\caption{Sketch of a multi-port system and time reversal. (a) Structure with $m$ ports. For incoming waves with unit amplitude at the $k$-th port, the output amplitude at each port is $S_{jk}$ with $j=1,2,\cdots,m$. The field induced by this input is ${\bf E}_k$.  (b) Effects of time-reversal operation. The directions of propagation of all waves are reversed and the amplitudes become the complex conjugates.}
\label{fig1}
\end{figure}

Consider a time-independent structure with multiple ports. The amplitudes of input and output waves are described by a scattering matrix or $S$~matrix. By convention, if the $k$-th port receives an input wave with unit amplitude while all other ports have zero input, the output amplitude at the $j$-th port is $S_{jk}$. The distribution of electric fields excited from port~$k$ is denoted by ${\bf E}_k$, as shown in Fig.~\ref{fig1}(a). In this paper, we work in frequency domain and the field amplitudes take complex values.
Time reversal reverses propagation of waves and conjugates the complex amplitudes, as depicted in Fig.~\ref{fig1}(b). If the system possesses time-reversal symmetry and linearity, the complex-conjugate amplitude ${\bf E}^*_k$ can be expressed as a linear combination of ${\bf E}_j$ by ${\bf E}_k^*=\sum_{j=1}^{m} S^*_{jk}{\bf E}_j$, where $m$ is the number of ports~\cite{Sounas2017}. A similar relation $(\hat{L} {\bf E}_k)^*=\sum_{j=1}^{m} S^*_{jk} \hat{L} {\bf E}_j$ holds for any real linear operator $\hat L$ acting on all fields, since $\hat L$ commutes with conjugation and the scalars $S_{jk}^*$. To formulate our results, let us introduce a column vector $\mathcal E$ composed of the $\hat L{\bf E}_j$, i.e., $\mathcal{E} = (\hat L{\bf E}_1,\hat L{\bf E}_2,\cdots\cdots,\hat L{\bf E}_m)^{\top}$, which thus satisfies
\begin{equation}\label{eq:ESE}
\mathcal{E}^* = S^\dagger\mathcal{E},
\end{equation}
with $*$ and $\dagger$ denoting complex conjugation and conjugate transposition, respectively. 
If each $\mathcal{E}_j=\hat L{\bf E}_j$ has multiple elements, $S$ here is understood as $S\otimes\mathbbm{1}$, with $\otimes$ and $\mathbbm{1}$ denoting the Kronecker product and the identity operator acting on each $\mathcal{E}_j$, respectively. For example, if $\mathcal{E}_j$ has $n$ components representing fields at different locations and/or in different directions, $\mathbbm{1}$ is understood as $\mathbbm{1}_n$; if $\mathcal{E}_j$ is a function, $\mathbbm{1}$ is understood as the identity operator in the function space. (More generally, each $\mathcal{E}_j$ may be an element in a Hilbert space with inner product of two vectors defined as $\mathcal{E}_j^\dagger\mathcal{E}_k$ for discrete cases or $\int \mathcal{E}_j^*\mathcal{E}_k$ for functions.) Our discussion hereafter assumes discrete cases with $n=1$ unless otherwise stated, but the extension to $n>1$ or functions is straightforward (e.g., matrices become block matrices).

Starting from the above relation and defining a matrix function
\begin{equation}\label{eq:g}
g_S(X)=\Re X+F_S\Re X{F_S}^\top+F_S\Im X-\Im X{F_S}^\top,
\end{equation}
with $F_S=\Im S(\mathbbm{1}_m+\Re S)^{-1}$ and $\top$ representing matrix transposition, our main result (proved in Appendix~\ref{app:rayleigh}) is bounds on the generalized Rayleigh quotient for chosen $m\times m$ Hermitian matrices $V$ and $W$ (depending on the quantity being bounded, as discussed below):
\begin{equation}\label{eq:Rayleigh-bounds}
\lambda_{\min}[G_S(V,W)]\le\frac{\mathcal{E}^\dagger W\mathcal{E}}{\mathcal{E}^\dagger V\mathcal{E}}
\le \lambda_{\max}[G_S(V,W)],
\end{equation}
where $G_S(V,W)=g_S(V)^{-1}g_S(W)$ is a matrix, $V$ is chosen so that $g_S(V)$ is positive-definite or negative-definite, $\lambda_{\min}$ and $\lambda_{\max}$ denote the minimum and maximum eigenvalues (guaranteed to be real), and $\mathcal{E}\ne0$ (but the fields involved may not necessarily be inside the scattering media or devices). We also assume that $\mathbbm{1}_m+\Re S$ is invertible, which is almost always satisfied (A remedy for singular cases is mentioned in Appendix~\ref{app:unitary-transform}.) The bounds are attained at $\mathcal{E}=(\mathbbm{1}+i{F_S}^\top)\psi$, where $\psi$ is the corresponding eigenvector of $G_S(V,W)$.
Various inequalities can be produced by Eq.~(\ref{eq:Rayleigh-bounds}) via different choices of~$\hat{L}$, $W$,~and~$V$. 

The most straightforward  application of Eq.~(\ref{eq:Rayleigh-bounds}) is to bound \emph{relative intensities} (or any linear combination thereof) of the fields (e.g.~at a single point, and/or a single direction, and/or volume-averaged, etcetera) excited from different ports.
In particular, with real diagonal matrices $V=\mathbbm{1}_m$ and $W={\rm diag}(w_1,\cdots,w_m)$, Eq.~(\ref{eq:Rayleigh-bounds}) becomes
\begin{equation}\label{eq:fraction-bounds}
\lambda_{\min}[G_S(V,W)]\le\sum_{k=1}^m w_k\eta_k\le \lambda_{\max}[G_S(V,W)],
\end{equation}
where $\eta_k$ is an ``intensity ratio'' defined by
\begin{equation}\label{eq:inten-frac}
\eta_k = \frac{\|\mathcal{E}_k\|^2}{\sum_{j=1}^m \|\mathcal{E}_j\|^2}.
\end{equation}
For a specified location $\bf r$, typical choices of each element of $\mathcal{E}$ (choices of $\hat{L}$) are a field component $\mathcal{E}_k={\bf E}_k({\bf r})\cdot\hat{n}$ in direction $\hat{n}$ or the entire field vector $\mathcal{E}_k={\bf E}_k({\bf r})$. (For choices like $\mathcal{E}_k={\bf E}_k({\bf r})$ that contain multiple components, similar to $S\otimes\mathbbm{1}_n$ above, the matrices $V$ and $W$ could be understood as $V\otimes\mathbbm{1}_n$ and $W\otimes\mathbbm{1}_n$.) One can also allow each element of $\mathcal{E}$ to involve multiple locations or even continuous volumes.
For example, consider the choice $\mathcal{E}_k=\sqrt{\epsilon}{\bf E}_k$ with $\epsilon$ being a real symmetric positive-definite permittivity tensor, where $\epsilon$ and $\bf E$ are functions of location. This choice, with the usual integral inner product~\cite{Joannopoulos2008}, yields the fraction of electric-field energy:
\begin{equation}\label{eq:energy-frac}
\eta_k = \frac{\int d^3{\bf x}\,{\bf E}_k({\bf x})\cdot{\bf D}_k({\bf x})}{\sum_{j=1}^m \int d^3{\bf x}\,{\bf E}_j({\bf x})\cdot{\bf D}_j({\bf x})},
\end{equation}
where ${\bf D}=\epsilon{\bf E}$ is the electric displacement field and $\bf x$ denotes the location. This fraction is also bounded by the $S$~matrix as in Eq.~(\ref{eq:fraction-bounds}).

If $G_S(V,W)$ happens to be zero, the inequalities in Eq.~(\ref{eq:Rayleigh-bounds}) reduce to the equalities:
\begin{equation}\label{eq:trs-equality}
\mathcal{E}^\dagger W\mathcal{E}=0,~~~~~~g_S(W)=0.
\end{equation}
An example of such a situation is presented in Sec.~\ref{sec:2port}.

\section{Numerical verification on multi-port structures}\label{sec:verification}
Here, we numerically validate Eq.~(\ref{eq:fraction-bounds}) in several example systems, with the choice $\mathcal{E}_k={\bf E}_k({\bf r})\cdot\hat{n}$, where the vector $\hat{n}$ is set as coordinate directions $\hat x$, $\hat y$, or $\hat z$. We focus on the trade-off between a pair of intensity fractions $\eta_j$ and $\eta_k$, where $j$ and $k$ label two selected ports. This corresponds to choosing $w_j,w_k\in\mathbb{R}$ while setting all other weights $w_{\ell \ne j,k}$ to zero. Eq.~(\ref{eq:fraction-bounds}) then becomes
\begin{equation}\label{eq:fraction-pair-bounds}
\lambda_{\min}(w_j,w_k)\le w_j\eta_j+w_k\eta_k\le \lambda_{\max}(w_j,w_k),
\end{equation}
where $\lambda_{\min}(w_j,w_k)$ is a function of $w_j$ and $w_k$ for a given $S$. In a multi-port structure with $m\ge3$, given $w_j$ and $w_k$ that are not simultaneously zero, Eq.~(\ref{eq:fraction-pair-bounds}) typically permits a ``linear strip'' lying between two parallel lines in $(\eta_j, \eta_k)$ space. By varying the ratio $w_j/w_k$, this strip changes (in both orientation and width), and the intersection of the strips for all $w_j/w_k$ yields the feasible region for $\eta_j$ and $\eta_k$.   For any given system, the main goal of the numerical validation is to see how tight the bounds are: how completely do the observed values of $(\eta_j, \eta_k)$ (e.g., for different $\bf{r}$ and $\hat{n}$) fill up the feasible region allowed by our bounds?

\begin{figure*}[ht]
\centering
\includegraphics[width=1\linewidth]{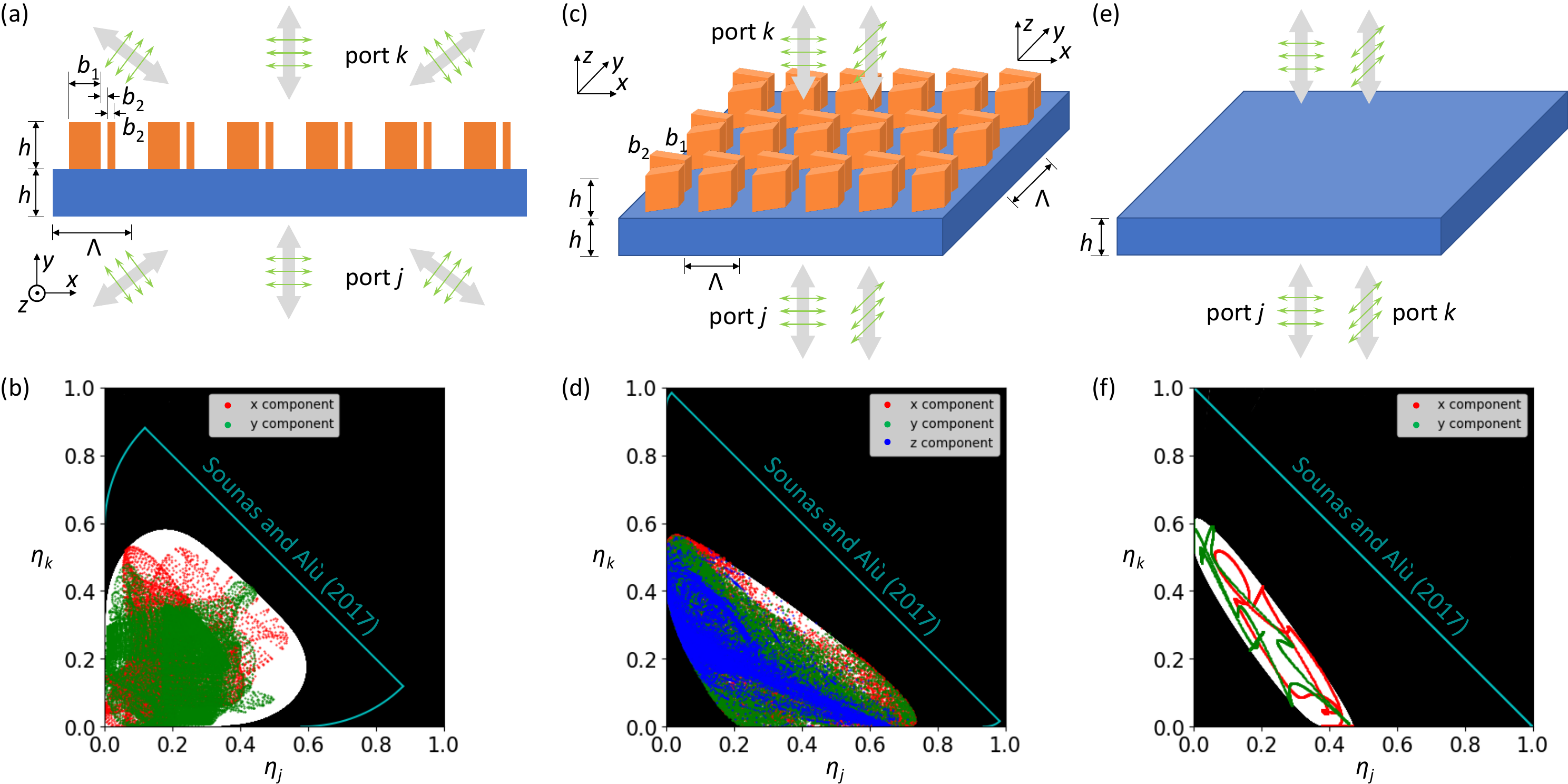}
\caption{Numerical validation based on multi-port systems. (a) Periodic grating composed of a substrate (silica, refractive index
$n = 1.45$) and ridges (silicon nitride, $n = 2.02$) with the minimum period $\Lambda$ in the $x$ direction and translational invariance in the $z$ (out-of-plane) direction. The height of the ridges and the thickness of the substrate are both $h=0.6\Lambda$. The widths of the ridges are $b_1=0.4\Lambda$ and $b_2=0.1\Lambda$. The width of the gap between the two ridges is also $b_2=0.1\Lambda$. The grating is illuminated by plane waves with in-plane polarization at the wavelength $0.8\Lambda$. The incoming plane waves are restricted to a few specific in-plane directions so that a six-port system is effectively formed. Our numerical validation focuses on the two ports labeled by $j$ and $k$.
(b) Feasible and infeasible regions predicted by time-reversal-symmetry bounds and data points from simulation for the system in (a). In the Cartesian coordinate system spanned by $\eta_j$ and $\eta_k$, the simulation data for $x$ and $y$ field components are represented by red and green dots, respectively. The white region is permitted by Eq.~(\ref{eq:fraction-bounds}) while the black region is forbidden. The cyan curve represents the bounds in Eq.~(\ref{eq:bounds2017}).
(c) Periodic grating composed of a substrate (silica) and rectangular pillars (silicon nitride). The structure forms a square lattice with a period $\Lambda$ in both $x$ and $y$ directions. The height of the pillars and the thickness of the substrate are both $h=0.75\Lambda$. The length and width of each pillar are $b_1=0.625\Lambda$ and $b_2=0.25\Lambda$. Each unit cell contains two pillars making an angle of 60$^\circ$.
The wavelength is $1.25\Lambda$. With incoming plane waves propagating in $\pm z$ directions and polarized in $x$ or $y$ directions, a four-port system is effectively formed. Our numerical validation focuses on the two ports labeled by $j$ and $k$.
(d) Feasible and infeasible regions predicted by time-reversal-symmetry bounds and data points from simulation for the system in (c). The simulation data for $x$, $y$, and $z$ field components are represented by red, green, and blue dots, respectively. The coordinate system, white and black regions, and the cyan curve have the same meaning as that in (b).
(e) Time-reversal-symmetric but nonreciprocal film suspended in vacuum with normal-incidence plane waves polarized in $x$ or $y$ directions. The film is made of a fictitious medium with a real but asymmetric permittivity tensor. The ratio of film thickness to wavelength is 1.56.
(f) Feasible and infeasible regions predicted by time-reversal-symmetry bounds and data points from simulation for the system in (e). The coordinate system, red and green dots, and white and black regions have the same meaning as that in (b). The cyan curve depicts the bound in the second inequality of Eq.~(\ref{eq:bounds2017}).
}
\label{fig2}
\end{figure*}

In this section, all media are dielectric with real permittivity, vacuum magnetic permeability, and zero magnetoelectric couplings.  We consider three examples, shown in Fig.~\ref{fig2} and discussed below.  Electromagnetic simulations were performed with a free-software implementation of the finite-difference time-domain (FDTD) method~\cite{Oskooi2010}, except where otherwise noted.   In each case, we show the feasible set predicted by our bounds as a white region, the infeasible set as a black region, and numerical observations as colored data points.  For comparison, we also plot the relevant bounds from the previous work~\cite{Sounas2017} as the cyan curves, where the curvilinear and straight segments correspond to
\begin{equation}\label{eq:bounds2017}
\eta_j+\eta_k-2|S_{jk}|\sqrt{\eta_j\eta_k}\le 1-|S_{jk}|^2,~~~~
\eta_j+\eta_k\le 1,
\end{equation}
respectively.
(For small $|S_{jk}|$, the former inequality approaches the latter, while the latter is immediately implied by the definition of the fractions in Eq.~(\ref{eq:inten-frac}), which sum up to one.)  In each example, we find that the numerical results lie within our predicted feasible set, and nearly saturate our constraints, whereas the bounds of Ref.~\citenum{Sounas2017} are much looser. The tightness of our inequalities~(\ref{eq:Rayleigh-bounds}), (\ref{eq:fraction-bounds}), and (\ref{eq:fraction-pair-bounds}) compared with the previous inequalities~(\ref{eq:bounds2017}) can be attributed to two reasons. First, our inequalities exploit the entire $S$ matrix instead of a single $S$ parameter. Second, the first inequality in Eq.~(\ref{eq:bounds2017}) was derived via the Cauchy–Schwarz inequality, which weakens the bounds, whereas the derivation of our inequalities, as detailed in Appendix~\ref{app:rayleigh}, does not involve a weakening step. In addition, the first inequality in Eq.~(\ref{eq:bounds2017}) relies on power conservation but our inequalities~(\ref{eq:Rayleigh-bounds}), (\ref{eq:fraction-bounds}), and (\ref{eq:fraction-pair-bounds}) do not, which implies that our inequalities are more widely applicable.

First, we simulate an asymmetric six-port system based on a periodic grating composed of a substrate (silica, refractive index $n=1.45$) and ridges (silicon nitride, $n=2.02$) in vacuum, as shown in Fig.~\ref{fig2}(a). The incident light is monochromatic plane waves in the $xy$ plane with the $H_z$ (in-plane $\bf{E}$ field) polarization. We consider fields satisfying periodic boundary conditions, which effectively restricts incoming and outgoing waves to be either normal incidence (perpendicular to the substrate) or be one of a finite set of diffracted orders~\cite{Joannopoulos2008}.  By choosing a wavelength in the range $\Lambda/2<\lambda<\Lambda$ for the minimum period $\Lambda$, there are only two possible diffracted beams on either side, effectively form a system with six ports as shown in Fig.~\ref{fig2}(a).
To validate Eq.~(\ref{eq:fraction-bounds}), the $S$~matrix and fields induced by a unit-amplitude input from each single port need to be computed. We evaluate the in-plane field intensities $|{\bf E}_\ell\cdot{\hat{n}}|^2$ for $\ell=1,\cdots,6$ and ${\hat{n}}=\hat x, \hat y$ in the grating, the substrate, and nearby free space.
Here, the two selected ports labeled by $j$ and $k$ are shown in Fig.~\ref{fig2}(a). The feasible region for $\eta_j$ and $\eta_k$, illustrated as the white area in Fig.~\ref{fig2}(b), is nearly filled by the dots, which represent the simulation data. The data points close to the diagonal correspond to the locations where fields can be excited strongly from both ports. In contrast, the data points in the vicinity of the origin come from locations that are nearly inaccessible from either of the two ports. In this example, the origin is included in the feasible region.
(As discussed in Appendix~\ref{app:null-intensity}, in an $m$-port system, time-reversal symmetry does not forbid intensity-like fractions in Eq.~(\ref{eq:inten-frac}) associated with at most $\lfloor (m-1)/2\rfloor$ ports from being zero simultaneously at certain locations.)

In the second example, we consider a 3d grating with two-dimensional periodicity along both $x$ and $y$ directions. The minimum period is the same for both directions and the media are the same as those in the previous example.
Let the incident light be monochromatic plane waves that propagate perpendicularly to the substrate surface and have a wavelength too large to allow diffracted orders. With both linear ($E_x$ and $E_y$) polarizations taken into account, such a system has four ports, each of which is assigned to either polarization on either side, as shown in Fig.~\ref{fig2}(c). Here, the V-shaped pattern in each unit cell mixes the two polarizations and hence makes this four-port system distinct from a trivial combination of two-port systems. We evaluate the field intensities $|{\bf E}_\ell\cdot{\hat{n}}|^2$ for $\ell=1,\cdots,4$ and ${\hat{n}}=\hat x,\hat y,\hat z$ at locations in the grating, the substrate, and nearby free space.
The observed intensity fractions $\eta_j$ and $\eta_k$ again nearly fill the feasible region calculated from the $S$~matrix, as shown in Fig.~\ref{fig2}(d). In contrast to the six-port system above, the origin $\eta_j = \eta_k = 0$ is excluded from the feasible region, as discussed in Appendix~\ref{app:null-intensity}. In other words, $\eta_j=\eta_k=0$ is forbidden at any location in this system.

In the third example, we consider a medium with a real asymmetric permittivity tensor, which breaks reciprocity but still preserves time-reversal symmetry~\cite{Buddhiraju2020}, in contrast to the previous systems that have both Lorentz reciprocity and time-reversal symmetry. A thin film made of such a medium is suspended in vacuum, as Fig.~\ref{fig2}(e) shows. This medium is assigned a relative permittivity
\begin{equation}
\epsilon=\begin{pmatrix}\epsilon_{xx}&\epsilon_{xy}&\epsilon_{xz}\\\epsilon_{yx}&\epsilon_{yy}&\epsilon_{yz}\\\epsilon_{zx}&\epsilon_{zy}&\epsilon_{zz}\end{pmatrix}=
\begin{pmatrix}2&1.5&0\\0.9&3&0\\0&0&1\end{pmatrix}.
\end{equation}
For normal-incidence plane waves, this system has four ports (two polarizations on each side), similar to the previous example. The $S$~matrix and fields here are computed via a transfer-matrix method~\cite{Mackay2020}.
Given two ports labeled by $j$ and $k$, the intensity fractions $\eta_j$ and $\eta_k$ evaluated in the film and nearby free space for both $x$ and $y$ components again satisfy (and nearly saturate) the time-reversal-symmetry bounds, as Fig.~\ref{fig2}(f) shows.  As in the previous example, the origin $\eta_j = \eta_k = 0$ is excluded by the bounds.

\section{Properties and corollaries}\label{sec:remarks}
In this section, we discuss some immediate corollaries of our time-reversal-symmetry bounds. Additional discussion of the properties of the bounds can be found in Appendices~\ref{app:unitary-transform} and \ref{app:sub-systems}.

\subsection{Bounds on other types of fields}
For fields that transform in the same way as $\bf E$ under time reversal, such as electric displacement fields, the bounds still apply. If a minus sign appears in addition to complex conjugate under time reversal, such as magnetizing fields and magnetic fields, either side of Eq.~(\ref{eq:ESE}) and hence $F_S$ in Eq.~(\ref{eq:g}) should be multiplied by $-1$ and becomes $-\Im S(\mathbbm{1}+\Re S)^{-1}$. If both $V$ and $W$ are real matrices, like those for intensity fractions in Eq.~(\ref{eq:fraction-bounds}), the bounds are the same for all these fields with or without the minus sign in $F_S$. However, some quantities, such as $\int d^3{\bf x}\,{\bf J}^*\cdot{\bf E}$ with a general complex time-invariant vector field $\bf J$, do not transform in either way. Nontrivial bounds on the generalized Rayleigh quotient for this integral may be beyond the reach of time-reversal symmetry alone, but related quantities can be bounded as described in Appendix~\ref{sec:complex-vector}.

\subsection{Bounds on emitted waves}\label{subsec:emit}
If a point dipole source couples with the output modes, the emitted-wave amplitudes at the ports are bounded in a similar way as Eq.~(\ref{eq:Rayleigh-bounds}) under time-reversal symmetry. (If the scattering device is a one- or two-dimensional periodic structure, a periodic array of point dipole sources are considered here. Possible extensions to Bloch-periodic and isolated sources in periodic structures are discussed in Sec.~\ref{sec:conclusion}.) To be specific, the amplitude $A_k$ of the emitted wave at the $k$-th port from a point dipole source with the direction $\hat{n}$ and location $\bf r$ is proportional to the field ${\bf E}_k({\bf r})\cdot{\hat{n}}$ evaluated in a complementary system whose $S$~matrix is $S^\top$, as detailed in Appendix~\ref{app:emit}. Consequently, the emitted power $P_k$ at the $k$-th port is proportional to $|{\bf E}_k({\bf r})\cdot{\hat{n}}|^2$ in the complementary system. Averaging over all directions $\hat{n}$, the mean emitted power is proportional to $|{\bf E}_k({\bf r})|^2$ in the complementary system.

Therefore, with $S$ replaced by $S^\top$ in Eq.~(\ref{eq:Rayleigh-bounds}), time-reversal symmetry constrains the amplitudes of emitted waves as 
\begin{equation}\label{eq:Rayleigh-bounds-emit}
\lambda_{\min}[G_{S^\top}(V,W)]\le\frac{\mathcal{A}^\dagger W\mathcal{A}}{\mathcal{A}^\dagger V\mathcal{A}}
\le \lambda_{\max}[G_{S^\top}(V,W)],
\end{equation}
where $\mathcal{A}=(A_1,\cdots,A_m)^\top$ is a column vector consisting of the emitted-wave amplitude $A_k$ at each port $k$. Similarly, the emitted power $P_k\propto|A_k|^2$ is constrained as
\begin{equation}\label{eq:fraction-bounds-emit}
\begin{aligned}
&\lambda_{\min}[G_{S^\top}(\mathbbm{1},{\rm diag}(w_1,\cdots,w_m))]\le\sum_{k=1}^m w_k\eta_k^{\rm emit}\\
&\le \lambda_{\max}[G_{S^\top}(\mathbbm{1},{\rm diag}(w_1,\cdots,w_m))],
\end{aligned}
\end{equation}
with the emitted-power fraction defined as
\begin{equation}
\eta_k^{\rm emit} = \frac{P_k}{\sum_{j=1}^m P_j}.
\end{equation}
A system satisfying Lorentz reciprocity implies $S=S^\top$, in which case the excited fields and the emitted-wave amplitudes are constrained by the same bounds. Among time-reversal symmetry, Lorentz reciprocity, and energy conservation, any two imply the third~\cite{Haus1984,Zhao2019,Buddhiraju2020}. Therefore, if both time-reversal symmetry and energy conservation are satisfied, the bounds on the excited fields and the emitted-wave amplitudes also coincide.

\section{Application to coupled resonance}\label{sec:tcmt}
In this section, we apply the time-reversal-symmetry bounds in Eqs.~(\ref{eq:Rayleigh-bounds}) and (\ref{eq:fraction-bounds}) to coupled-resonant systems described by temporal coupled-mode theory (TCMT), valid for long-lifetime resonances~\cite{Fan2003}. In previous work, with time-reversal symmetry and energy conservation satisfied simultaneously, resonant decay coefficients were identical to resonant coupling coefficients, both of which were subject to the same bounds~\cite{Wang2013,Wang2024}. Here, with only time-reversal symmetry being assumed, resonant decay and coupling rates are not necessarily identical and can be bounded separately.

Let us consider a single-mode optical resonance coupled with $m$ ports. In TCMT~\cite{Fan2003}, the dynamics of the amplitude $a$ of the resonant mode is described by
\begin{equation}\label{eq:tcm}
\frac{da}{dt} = \left(i\omega_0-\gamma\right)a+\bm{\kappa}^\top{\bf{s}}_+,~~~~
{\bf{s}}_- = C{\bf{s}}_++a\bf{d},
\end{equation}
where $\omega_0$ denotes the resonant frequency, $\gamma$ denotes the resonant decay rate (with its inverse being the lifetime of resonance), and $t$ denotes time. The incoming and outgoing wave amplitudes are denoted by $\bf{s}_+$ and $\bf{s}_-$, respectively. The coupling coefficients between $\bf{s}_+$ and the resonant mode are denoted by $\bm{\kappa}$. The coupling coefficients between $\bf{s}_-$ and the resonant mode, namely the decay coefficients, are denoted by $\bf{d}$. The direct pathway without resonance is described by the scattering matrix $C$. Here, each of ${\bf{s}}_+$, ${\bf{s}}_-$, $\bm{\kappa}$, and $\bf{d}$ is a column vector with $m$ elements, and $C$ is an $m\times m$ matrix.

If the system obeys time-reversal symmetry, we prove in Appendix~\ref{app:TCMT} that the resonant coupling and decay coefficients are bounded by a very similar Rayleigh-quotient relation:
\begin{equation}\label{eq:Rayleigh-bounds-tcmt}
\begin{aligned}
&\lambda_{\min}[G_C(V,W)]\le\frac{\bm{\kappa}^\dagger W\bm{\kappa}}{\bm{\kappa}^\dagger V\bm{\kappa}}
\le \lambda_{\max}[G_C(V,W)],\\
&\lambda_{\min}[G_{C^\top}(V,W)]\le\frac{{\bf{d}}^\dagger W\bf{d}}{{\bf d}^\dagger V\bf{d}}
\le \lambda_{\max}[G_{C^\top}(V,W)],
\end{aligned}
\end{equation}
with $G_C(V,W)=g_C(V)^{-1}g_C(W)$ and $G_{C^\top}(V,W)=g_{C^\top}(V)^{-1}g_{C^\top}(W)$, where $g_C$ and $g_{C^\top}$ are defined in the same fashion as in Eq.~(\ref{eq:g}). In particular, similar to Eqs.~(\ref{eq:fraction-bounds}) and (\ref{eq:inten-frac}), with $V=\mathbbm{1}_m$ and a real diagonal matrix $W={\rm diag}(w_1,\cdots,w_m)$, Eq.~(\ref{eq:Rayleigh-bounds-tcmt}) yields inequalities for the resonant coupling rates $|\kappa_k|^2$ and decay rates $|d_k|^2$:
\begin{equation}\label{eq:rate-bounds-tcmt}
\begin{aligned}
&\lambda_{\min}[G_C(V,W)]\le\frac{\sum_{k=1}^m w_k|\kappa_k|^2}{\sum_{k=1}^m|\kappa_k|^2}
\le \lambda_{\max}[G_C(V,W)],\\
&\lambda_{\min}[G_{C^\top}(V,W)]\le\frac{\sum_{k=1}^m w_k|d_k|^2}{\sum_{k=1}^m|d_k|^2}
\le \lambda_{\max}[G_{C^\top}(V,W)].
\end{aligned}
\end{equation}

If the system has Lorentz reciprocity, $C$ is symmetric and $\bm{\kappa}=\bm{d}$. With both time-reversal symmetry and Lorentz reciprocity, ${\bm{d}}$ and $\bm{\kappa}$ obey the same bounds and the denominators in Eq.~(\ref{eq:rate-bounds-tcmt}) can be explicitly written as $\bm{d}^\dagger\bm{d}=\bm{\kappa}^\dagger\bm{\kappa}=2\gamma$~\cite{Fan2003}. With the resonant coupling and decay rate $\gamma_k$ for each port given as $|d_k|^2=|\kappa_k|^2=2\gamma_k$, Eq.~(\ref{eq:rate-bounds-tcmt}) reduces to
\begin{equation}\label{eq:rate-bounds-tcmt-recipr}
\lambda_{\min}[G_C(V,W)]\le\frac{\sum_{k=1}^m w_k\gamma_k}{\gamma}
\le \lambda_{\max}[G_C(V,W)].
\end{equation}

For multiple resonant modes coupled with each other and with multiple ports~\cite{Suh2004}, the time-reversal-symmetry bounds in Eqs.~(\ref{eq:Rayleigh-bounds-tcmt}) and (\ref{eq:rate-bounds-tcmt}) apply separately to each resonant mode in the context of TCMT.

\section{Two-port structure}\label{sec:2port}
\subsection{Amplitude-phase relation in a two-port structure}
In this section, we study a two-port structure possessing time-reversal symmetry in both the general case with $\mathcal{E}_k=\hat L{\bf E}_k$ and coupled-resonant systems described by TCMT. The results here can stem from the time-reversal-symmetry equality~(\ref{eq:trs-equality}) as sketched below, and can also be derived directly starting from Eq.~(\ref{eq:ESE}) without resorting to Eq.~(\ref{eq:trs-equality}). One can also obtain the results in this section using more complicated approaches.

The general form of a two-port $S$~matrix possessing time-reversal symmetry is
\begin{equation}\label{eq:2port-S}
S = e^{i\zeta}\begin{pmatrix}
e^{i\phi}\sqrt{1-t_1t_2} & t_1 \\
t_2 & -e^{-i\phi}\sqrt{1-t_1t_2}
\end{pmatrix},
\end{equation}
with $\zeta,\phi,t_1,t_2\in\mathbb{R}$, and $t_1t_2\le1$. Solving $g_S(W)=0$ yields
\begin{equation}
W \propto \begin{pmatrix}
-t_1 & \sqrt{1-t_1t_2}e^{i\phi} \\
\sqrt{1-t_1t_2}e^{-i\phi} & t_2
\end{pmatrix}.
\end{equation}
According to Eq.~(\ref{eq:trs-equality}), $W$ then satisfies $\mathcal{E}^\dagger W\mathcal{E}=0$, which can be rearranged algebraically to:
\begin{equation}\label{eq:2port-trs-circle}
\frac{|\mathcal{E}_2|^2}{|\mathcal{E}_1|^2}+\frac{2\sqrt{1-t_1t_2}\cos(\theta+\phi)}{t_2}\frac{|\mathcal{E}_2|}{|\mathcal{E}_1|}=\frac{t_1}{t_2},
\end{equation}
where $\theta$ denotes the phase difference $\theta=\arg(\mathcal{E}_1^*\mathcal{E}_2)$ and we assume $t_2\ne0$. If $t_1\ne0$ but $t_2=0$, one can rewrite Eq.~(\ref{eq:2port-trs-circle}) or relabel the two ports; if $t_1=t_2=0$, $\mathcal{E}^\dagger W\mathcal{E}=0$ reduces to $|\mathcal{E}_1||\mathcal{E}_2|\cos(\theta+\phi)=0$. The equality~(\ref{eq:2port-trs-circle}) not only reveals the strong dependence between the intensity ratio and the phase difference, but also yields
\begin{equation}\label{eq:2port-trs-bounds}
\frac{|1-\sqrt{1-t_1t_2}|}{|t_2|}\le\frac{|\mathcal{E}_2|}{|\mathcal{E}_1|}\le \frac{1+\sqrt{1-t_1t_2}}{|t_2|},
\end{equation}
which can be obtained more straightforwardly by setting $w_1=1$ and $w_2=-1$ in Eq.~(\ref{eq:fraction-bounds}) with $S$ in Eq.~(\ref{eq:2port-S}). According to Eqs.~(\ref{eq:Rayleigh-bounds-emit}) and (\ref{eq:fraction-bounds-emit}), the emitted-wave amplitudes and radiated power at the ports from an internal point dipole source are constrained similarly as Eqs.~(\ref{eq:2port-trs-circle}) and (\ref{eq:2port-trs-bounds}):
\begin{equation}
\frac{|A_2|^2}{|A_1|^2}+\frac{2\sqrt{1-t_1t_2}\cos(\theta+\phi)}{t_1}\frac{|A_2|}{|A_1|}=\frac{t_2}{t_1},
\end{equation}
\begin{equation}
\frac{|1-\sqrt{1-t_1t_2}|}{|t_1|}\le\sqrt{\frac{P_2}{P_1}}\le \frac{1+\sqrt{1-t_1t_2}}{|t_1|}.
\end{equation}

\begin{figure}[ht]
\centering
\includegraphics[width=0.86\linewidth]{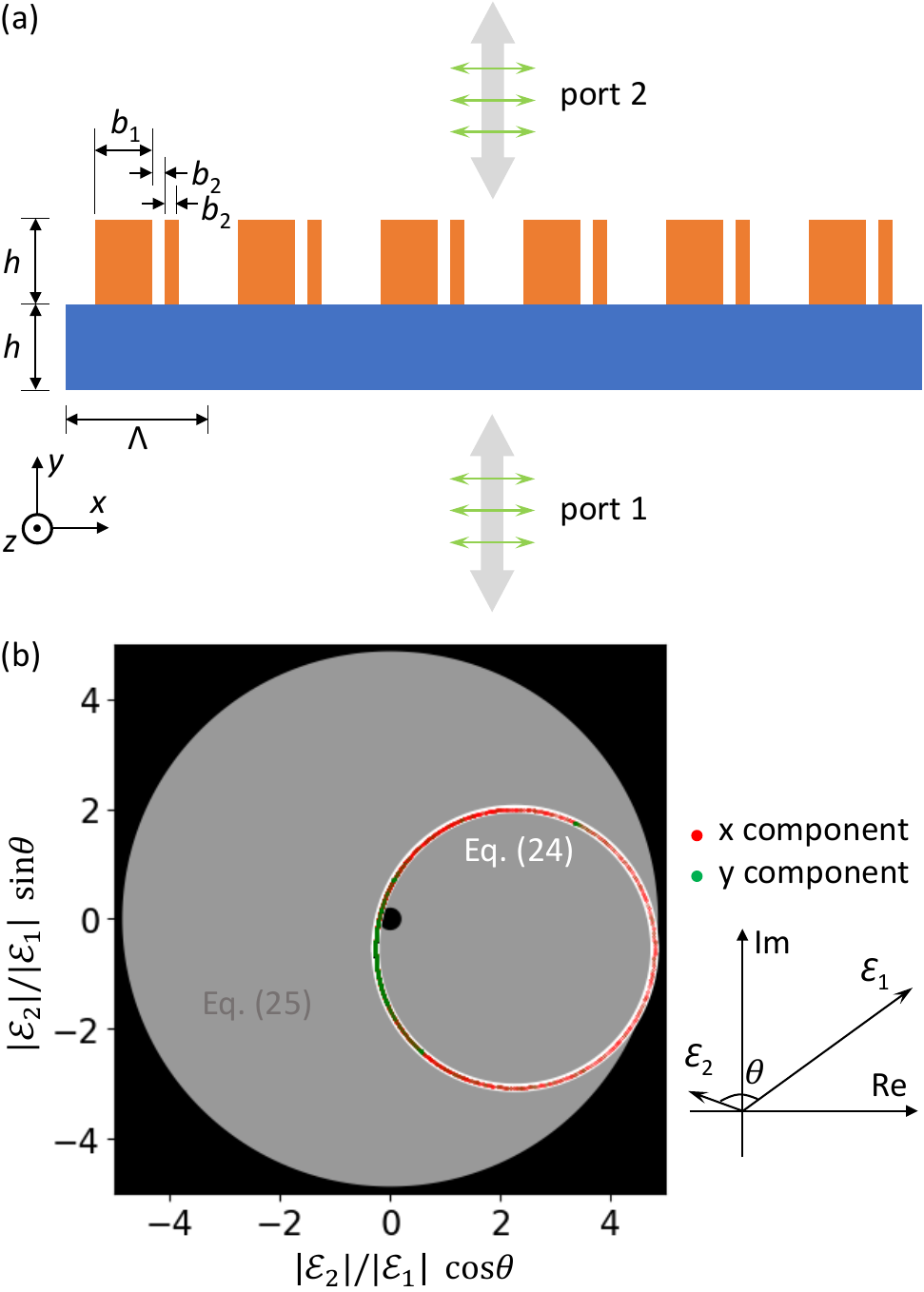}
\caption{Numerical validation of the amplitude-phase relation on a two-port system. (a) Periodic grating same as that in Fig.~\ref{fig2}(a). The wavelength is $1.2\Lambda$, so that the period is subwavelength, which allows only zeroth-order diffraction in transmission and reflection.
(b) Feasible and infeasible regions predicted by time-reversal-symmetry bounds and data points from simulation for the system in (a). We use a polar-coordinate system with the polar axis and polar angle being $|\mathcal{E}_2|/|\mathcal{E}_1|$ and $\theta=\arg(\mathcal{E}_1^*\mathcal{E}_2)$, respectively. The gray and black areas are allowed and disallowed by Eq.~(\ref{eq:2port-trs-recipr-bounds}), respectively. On top of the gray area, only the white circle satisfies Eq.~(\ref{eq:2port-trs-recipr-circle}) and thus is permitted, which is occupied by the red and green dots corresponding to $x$ and $y$ field components from numerical simulation.
}
\label{fig3}
\end{figure}

If the system additionally has Lorentz reciprocity, yielding $t_1=t_2$, Eqs.~(\ref{eq:2port-trs-circle}) and (\ref{eq:2port-trs-bounds}) are simplified as
\begin{equation}\label{eq:2port-trs-recipr-circle}
\frac{|\mathcal{E}_2|^2}{|\mathcal{E}_1|^2}+\frac{2r\cos(\theta+\phi)}{t}\frac{|\mathcal{E}_2|}{|\mathcal{E}_1|}=1,
\end{equation}
\begin{equation}\label{eq:2port-trs-recipr-bounds}
\frac{1-r}{1+r}\le\frac{|\mathcal{E}_2|^2}{|\mathcal{E}_1|^2}\le \frac{1+r}{1-r},
\end{equation}
with $t=t_1=t_2$ and $r=\sqrt{1-t^2}$. Due to reciprocity, the emitted-wave amplitude $A_k$ and power $P_k$ of this system obey the same relations as $\mathcal{E}_k$ and $|\mathcal{E}_k|^2$, respectively.
In a polar-coordinate system with $|\mathcal{E}_2|/|\mathcal{E}_1|$ and $\theta=\arg(\mathcal{E}_1^*\mathcal{E}_2)$ being the polar axis and polar angle, respectively, each of Eqs.~(\ref{eq:2port-trs-circle}) and (\ref{eq:2port-trs-recipr-circle}) represents a circle, while each of Eqs.~(\ref{eq:2port-trs-bounds}) and (\ref{eq:2port-trs-recipr-bounds}) represents an annulus, as exemplified in Fig.~\ref{fig3}(b). 

For a two-port optical resonator with time-reversal symmetry, with the direct-pathway scattering matrix $C$ taking the general form as Eq.~(\ref{eq:2port-S}), the resonant coupling coefficients $\kappa_1$ and $\kappa_2$ satisfy
\begin{equation}\label{eq:tcmt-2ports-special-circle-k}
\frac{|\kappa_2|^2}{|\kappa_1|^2}+\frac{2\sqrt{1-t_1t_2}\cos(\theta+\phi)}{t_2}\frac{|\kappa_2|}{|\kappa_1|}=\frac{t_1}{t_2},
\end{equation}
with $\theta=\arg(\kappa_1^*\kappa_2)$. This equality yields
\begin{equation}\label{eq:tcmt-2port-trs-bounds-k}
\frac{|1-\sqrt{1-t_1t_2}|}{|t_2|}\le\frac{|\kappa_2|}{|\kappa_1|}\le \frac{1+\sqrt{1-t_1t_2}}{|t_2|}.
\end{equation}
Likewise, the resonant decay $d_1$ and $d_2$ satisfy
\begin{equation}\label{eq:tcmt-2ports-special-circle-d}
\frac{|d_2|^2}{|d_1|^2}+\frac{2\sqrt{1-t_1t_2}\cos(\theta+\phi)}{t_1}\frac{|d_2|}{|d_1|}=\frac{t_2}{t_1},
\end{equation}
with $\theta=\arg(d_1^*d_2)$. This equality yields
\begin{equation}\label{eq:tcmt-2port-trs-bounds-d}
\frac{|1-\sqrt{1-t_1t_2}|}{|t_1|}\le\frac{|d_2|}{|d_1|}\le \frac{1+\sqrt{1-t_1t_2}}{|t_1|}.
\end{equation}
Both Eqs.~(\ref{eq:tcmt-2port-trs-bounds-k}) and (\ref{eq:tcmt-2port-trs-bounds-d}) can also be obtained from Eq.~(\ref{eq:rate-bounds-tcmt}).
If the system also has Lorentz reciprocity that allows $t_1=t_2$, $\kappa_1=d_1$, and $\kappa_2=d_2$, Eqs.~(\ref{eq:tcmt-2ports-special-circle-k})-(\ref{eq:tcmt-2port-trs-bounds-d}) can all be simplified. In particular, Eqs.~(\ref{eq:tcmt-2port-trs-bounds-k}) and (\ref{eq:tcmt-2port-trs-bounds-d}) become
\begin{equation}\label{eq:tcmt-2ports-ratio-bounds}
\frac{1-r}{1+r}\le\frac{\gamma_2}{\gamma_1}\le \frac{1+r}{1-r},
\end{equation}
with $\gamma_k=|\kappa_k|^2=|d_k|^2$ representing the resonant coupling and decay rates. The same inequalities as Eq.~(\ref{eq:tcmt-2ports-ratio-bounds}) were also derived in Ref.~\cite{Wang2013}.

\subsection{Numerical verification on a two-port structure}
We simulate an asymmetric two-port system based on a periodic grating composed of a substrate (silica, refractive index $n=1.45$) and ridges (silicon nitride, $n=2.02$) in vacuum, as depicted in Fig.~\ref{fig3}(a). The incident light is monochromatic plane waves that propagate perpendicularly to the substrate, with in-plane polarization (i.e., magnetic field parallel to the ridges) and a wavelength too large to support diffracted orders, yielding a two-port system.
Given $r$ in Eq.~(\ref{eq:2port-trs-recipr-bounds}), the allowed range of the intensity ratio can be obtained. In a polar-coordinate system with the polar axis $|\mathcal{E}_2|/|\mathcal{E}_1|$ and polar angle $\theta=\arg(\mathcal{E}_1^*\mathcal{E}_2)$, Eq.~(\ref{eq:2port-trs-recipr-bounds}) permits only the gray annulus in Fig.~\ref{fig3}(b). Given also $\phi$, the feasible circle in Eq.~(\ref{eq:2port-trs-recipr-circle}) can be determined, as shown by the white circle on top the gray area in Fig.~\ref{fig3}(b). To test this equality, fields induced by a unit-amplitude input wave from each port were computed by an FDTD method~\cite{Oskooi2010}. We evaluated the in-plane field components ${\bf E}_k\cdot{\hat x}$ and ${\bf E}_k\cdot{\hat y}$ for $k=1,2$ in the grating, the substrate, and nearby free space. As illustrated by red ($x$ component) and green ($y$ component) dots that overlap the feasible circle in Fig.~\ref{fig3}(b), the observed fields indeed satisfy Eqs.~(\ref{eq:2port-trs-recipr-circle}) and fill up the feasible circle.

\section{Application to lossy media}\label{sec:absorption}
As time-reversal symmetry requires permittivity and permeability to be real, their imaginary parts, which are responsible for loss, must be negligible at the frequency of interest for our bounds in the preceding sections to hold. However, when the loss is small but not negligible, it turns out that the time-reversal-symmetry bounds can still impose constraints on some loss-induced features.

Let us set $\mathcal{E}_k=\sqrt{\sigma}{\bf E}_k$ with $\sigma=\omega\Im\epsilon$. Here, we assume that all media in the system have symmetric $\Re\epsilon$, real and symmetric magnetic permeability tensors, and zero magnetoelectric couplings; wherever $\sigma$ is nonzero, we assume it to be isotropic and positive. Therefore, the corresponding lossless system with $\sigma=0$ satisfies time-reversal symmetry and does not have gain or loss, while time-reversal symmetry breaking and power dissipation are introduced by $\sigma > 0$.
One can define the fraction of power dissipation as
\begin{equation}\label{eq:dissipate-frac}
\eta_k = \frac{\int d^3{\bf x}\,\sigma({\bf x})|{\bf E}_k({\bf x})|^2}{\sum_{j=1}^m \int d^3{\bf x}\,\sigma({\bf x})|{\bf E}_j({\bf x})|^2},
\end{equation}
where the domain of integration includes all regions with nonzero $\sigma$. On the other hand, when only the $k$-th port is excited by unit-amplitude input, the time-average dissipated power is related to the non-unitarity of $S$~by
\begin{equation}\label{eq:dissipate-S}
\int d^3{\bf x}\,\sigma({\bf x})|{\bf E}_k({\bf x})|^2\propto 1-\sum_{j=1}^m |S_{(\sigma)jk}|^2,
\end{equation}
where $S_{(\sigma)}$ denotes the $S$~matrix in the presence of nonzero $\sigma$.
Therefore, the fraction of dissipated power in Eq.~(\ref{eq:dissipate-frac}) associated with each port and the weighted sum can be written as
\begin{equation}\label{eq:dissipate-frac-S}
\begin{aligned}
&\eta_k = \frac{1-\sum_{j=1}^m |S_{(\sigma)jk}|^2}{\tr[\mathbbm{1}_m-{S_{(\sigma)}}^\dagger S_{(\sigma)}]},\\
&\sum_{k=1}^m w_k\eta_k=\frac{\tr\left\{W\left[\mathbbm{1}_m-{S_{(\sigma)}}^\dagger S_{(\sigma)}\right]\right\}}{\tr[\mathbbm{1}_m-{S_{(\sigma)}}^\dagger S_{(\sigma)}]},
\end{aligned}
\end{equation}
with $W={\rm diag}(w_1,\cdots,w_m)$ being a real diagonal matrix. For small loss, now let us approximate all ${\bf E}_k({\bf x})$ terms to zero-th order in $\sigma$ by the corresponding fields in the lossless ($\sigma = 0$) system.
This introduces an error in ${\bf E}_k({\bf x})$ that is first-order in $\sigma$, an error in Eq.~(\ref{eq:dissipate-S}) that is second-order in $\sigma$, and an error in the ratio of Eq.~(\ref{eq:dissipate-frac}) that is first-order in $\sigma$.  Consequently, if we bound Eq.~(\ref{eq:dissipate-frac} for the $\sigma = 0$ fields using Eq.~(\ref{eq:fraction-bounds}), which is applicable to the lossless fields, the results will also apply to the non-unitarity of $S$ in Eq.~(\ref{eq:dissipate-frac-S}) up to a first-order correction $O(\sigma)$:
\begin{equation}\label{eq:dissipate-fraction-bounds}
\begin{aligned}
\lambda_{\min}\left[G_{S_{(0)}}(V,W)\right]&\le\frac{\tr\left\{W\left[\mathbbm{1}_m-{S_{(\sigma)}}^\dagger S_{(\sigma)}\right]\right\}}{\tr[\mathbbm{1}_m-{S_{(\sigma)}}^\dagger S_{(\sigma)}]}+O(\sigma)\\
&\le \lambda_{\max}\left[G_{S_{(0)}}(V,W)\right],
\end{aligned}
\end{equation}
with $V=\mathbbm{1}_m$, $W={\rm diag}(w_1,\cdots,w_m)$, and $S_{(0)}$ representing the $S$~matrix of the lossless background system.

As an example, let us apply Eq.~(\ref{eq:dissipate-fraction-bounds}) to a two-port structure with loss, and investigate its accuracy for different magnitudes of $\sigma$. In a two-port structure with time-reversal symmetry, the reflectance is the same on both sides, as manifested by $|S_{11}|^2=|S_{22}|^2$ in Eq.~(\ref{eq:2port-S}). In the presence of lossy media, $|S_{11}|^2\ne|S_{22}|^2$ is possible, but the extent of such loss-induced asymmetry in reflectance is constrained (approximately) by our bounds. Here, we consider a reciprocal two-port system with $t_1=t_2$ in Eq.~(\ref{eq:2port-S}) and set $w_1=1, w_2=-1$ in Eq.~(\ref{eq:dissipate-fraction-bounds}). The asymmetry in reflectance normalized by total loss can then be bounded as
\begin{equation}\label{eq:dissipate-2port-bounds}
-r_{(0)}\le\frac{|S_{(\sigma)11}|^2-|S_{(\sigma)22}|^2}{2-\sum_{j=1}^2\sum_{k=1}^2|S_{(\sigma)jk}|^2}+O(\sigma)\le r_{(0)},
\end{equation}
where $r_{(0)}=\sqrt{1-t_1^2}=\sqrt{1-t_2^2}$ is the magnitude of the reflection amplitude in the lossless background system.
These inequalities align with intuition: in a structure with high transparency (small $r_{(0)}$), there can be little asymmetry in the lossy reflectances $|S_{(\sigma)11}|^2$ and $|S_{(\sigma)22}|^2$ because light from both sides passes through the structure and experiences similar loss; in contrast, for a structure with high reflectance, the field penetration (hence loss) can be very different for light incident from the two sides.

\begin{figure}[ht]
\centering
\includegraphics[width=1\linewidth]{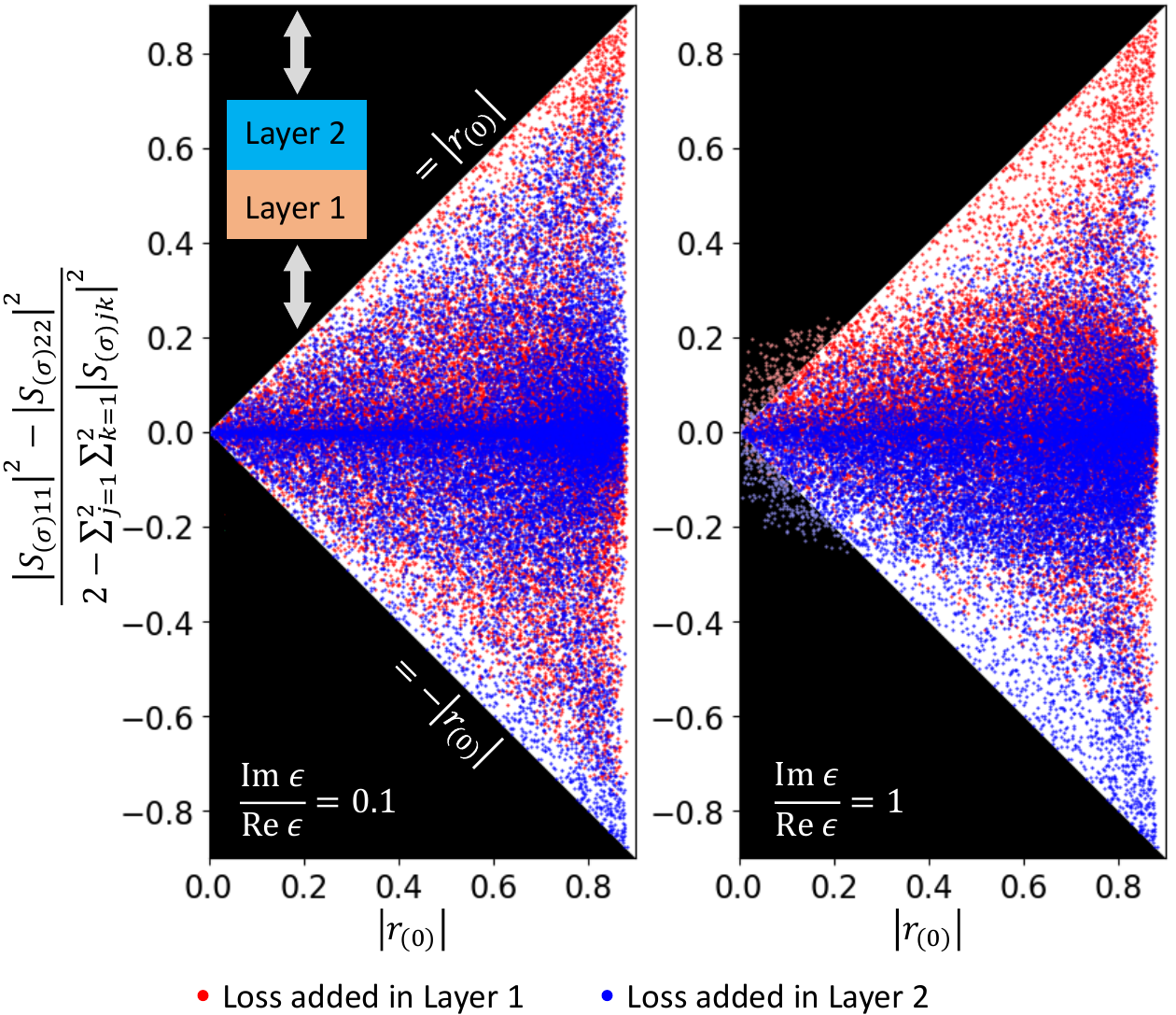}
\caption{Numerical validation of approximate (first-order) bounds on loss-induced asymmetric reflection on a two-port system. As shown in the inset, the structure is a double-layer film. The medium in each layer has homogeneous isotropic permittivity, vacuum permeability, and zero magnetoelectric coupling. The thickness and permittivity of each layer are picked randomly. These quantities of each layer are picked independently from those of the other layer. For a wavelength~$\lambda$ and relative permittivity $\epsilon$, the thickness of this layer is chosen in the interval $[0,\lambda/\sqrt{\Re\epsilon})$ and $\Re\epsilon/\epsilon_0$ is in $[1,16)$. When loss is added, the ratio $\Im\epsilon/\Re\epsilon$ is fixed as labeled in the lower-left corner of each figure. We always add loss in one layer while keeping the other lossless, with the choice of lossy layer indicated by red or blue dots. For loss added in each layer, 20000 random cases are generated. The white and black triangular regions are permitted and forbidden by Eq.~(\ref{eq:dissipate-2port-bounds}), respectively. The data points from simulation almost all lie in the white feasible region in (a), with a relative small dissipation; but in (b) with large dissipation, noticeable violations (dots in the black regions) are observed.
}
\label{fig4}
\end{figure}

We numerically verify Eq.~(\ref{eq:dissipate-2port-bounds}) for a two-layer film, as shown in the inset of Fig.~(\ref{fig4}). The incident light is monochromatic plane waves that propagate perpendicularly to the film, so that a two-port system is formed. The thicknesses and permittivities for the two layers in the film are selected randomly, and nonzero $\Im\epsilon$ is introduced in only one of the layers for any given structure. The $S$~matrix and fields here are computed via a transfer-matrix method~\cite{Mackay2020}.  The asymmetry in reflectance normalized by total loss very closely corresponds to the $\pm r_{(0)}$ bounds if $\Im\epsilon$ is relatively small, as the left panel of Fig.~\ref{fig4} shows. If $\Im\epsilon$ is not small enough compared to $\Re\epsilon$, noticeable violations can occur, as the right panel illustrates, but these violations of our approximate constraints are surprisingly rare even for $\Im\epsilon = \Re\epsilon$.

\section{Conclusion}\label{sec:conclusion}
In this work, we have formulated time-reversal-symmetry bounds on internal fields in linear electromagnetic systems in terms of the $S$~matrix and verified the bounds numerically. There are a number of potential extensions of this approach. Since the premises in this work are time-reversal symmetry and linearity, a natural question is how to make corrections if either or both premises are violated to a small but non-negligible extent. The bounds hold approximately if nonlinearity is negligibly small (at the powers and frequencies of interest), and could also be applied to linearizations of small perturbations around a nonlinear steady state. However, the treatment of larger nonlinearity, such as by a next-order correction, remains to be solved. Likewise, the bounds can hold to the extent that material loss is negligible, and we also showed in Sec.~\ref{sec:absorption} that first-order extensions to lossy systems can work surprisingly well to predict loss-induced changes in~$S$. Approximate extensions of the field-intensity bounds themselves for non-negligible loss seems an interesting open question.
Our approach may be potentially applied to estimate intensity-dependent behaviors in multi-port structures, e.g., the extent of nonreciprocity that relies on intensity-dependent permittivity~\cite{Sounas2018,Cotrufo2021a,Cotrufo2021b,Cotrufo2023,Wang2023} or the field overlap with a desired field pattern at different excitation arrangements~\cite{Craxton2015,Kirkwood2013}. For periodic systems, we have derived our bounds for periodic waves and/or point dipole sources, but it would be interesting to extend them to Bloch-periodic boundary conditions~\cite{Joannopoulos2008}, which poses a challenge because such boundary conditions break time-reversal symmetry; isolated point sources could then be analyzed via superpositions of Bloch-periodic point sources~\cite{Capolino2007}.
Finally, although our bounds in this work are derived for electromagnetic waves, similar results should apply to other physical contexts, such as acoustics and quantum mechanics, in which similar time-reversal principles appear~\cite{Fink2000,Sakurai2020}.

For experimental systems in which one can easily measure the $S$ matrix but not the internal fields within the structure, our bounds allow one to deduce information about this previously inaccessible regime from only exterior measurements.   For computational systems in which one is designing an asymmetric response to emission or excitation, perhaps by inverse design~\cite{Molesky2018,Christiansen2021,Kang2024}, theoretical bounds such as ours provide a measure of global optimality to help assess the local optima generated by non-convex optimization.

\begin{acknowledgments}
This work was supported in part by the US Army Research Office through the Institute for Soldier Nanotechnologies (Award No. W911NF-23-2-0121) and by the Simons Foundation through the Simons Collaboration on Extreme Wave Phenomena Based on Symmetries. The authors are grateful to Dimitrios~L.~Sounas and Andrea~Al{\`{u}} for helpful discussions.
\end{acknowledgments}



\appendix
\section{Bounds on Rayleigh quotients}\label{app:rayleigh}
If the vector $\mathcal{E}$ were not constrained by time-reversal symmetry but allowed to take any nonzero values, the generalized Rayleigh quotient in Eq.~(\ref{eq:Rayleigh-bounds}) would be bounded trivially by the minimum and maximum eigenvalues of the matrix $V^{-1}W$. Time-reversal symmetry constrains $\mathcal{E}$ as Eq.~(\ref{eq:ESE}), which shrinks the attainable range of the generalized Rayleigh quotient. To prove the nontrivial bounds in Eq.~(\ref{eq:Rayleigh-bounds}), instead of using the complex matrices $V,W$ and constrained complex vector $\mathcal{E}$ directly, one can recast the original quotient as an equivalent one with real matrices and an unconstrained real vector, which can be found via first expanding the field vector $\mathcal{E}$ and the $S$~matrix into their real and imaginary parts so that Eq.~(\ref{eq:ESE}) becomes
\begin{equation}\label{eq:ESE-real-imag}
\Re\mathcal{E}-i\Im\mathcal{E}=(\Re S^\top-i\Im S^\top)(\Re\mathcal{E}+i\Im\mathcal{E}).
\end{equation}
Equating the imaginary parts on both sides, one can obtain
\begin{equation}\label{eq:ESE-split}
(\mathbbm{1}_m+\Re S^\top)\Im\mathcal{E}=\Im S^\top\Re\mathcal{E},
\end{equation}
which is equivalent to
\begin{equation}\label{eq:ESE-Im-Re}
\Im\mathcal{E}={F_S}^\top\Re\mathcal{E}
\end{equation}
with $F_S=\Im S(\mathbbm{1}_m+\Re S)^{-1}$ if $\mathbbm{1}_m+\Re S$ is nonsingular, which is assumed hereafter; in the unlikely event that it is singular, this can be corrected by a unitary change of basis as discussed in Appendix~\ref{app:unitary-transform}. Hence $\Re\mathcal{E}$ can be chosen as an unconstrained real vector, which determines $\Im\mathcal{E}$ via Eq.~(\ref{eq:ESE-Im-Re}). (One can also choose $\Im\mathcal{E}$ as the unconstrained real vector that determined $\Re\mathcal{E}$, and will finally reach the same bounds.)

Expanding a Hermitian matrix $X$ and the vector $\mathcal{E}$ into their real and imaginary parts and employing Eq.~(\ref{eq:ESE-Im-Re}), one can obtain
\begin{equation}\label{eq:exe}
\begin{aligned}
\mathcal{E}^\dagger X\mathcal{E}=(\Re\mathcal{E})^\top g_S(X)\Re\mathcal{E},
\end{aligned}
\end{equation}
where $g_S(X)$, defined in Eq.~(\ref{eq:g}), is a function of $X$ given $S$ and always yields a real symmetric matrix from a Hermitian~$X$. For Hermitian matrices $V$ and $W$, the generalized Rayleigh quotient restricted by Eq.~(\ref{eq:ESE-split}) can be written as
\begin{equation}\label{eq:real-rayleigh}
\frac{\mathcal{E}^\dagger W\mathcal{E}}{\mathcal{E}^\dagger V\mathcal{E}}
= \frac{(\Re\mathcal{E})^\top g_S(W)(\Re\mathcal{E})}{(\Re\mathcal{E})^\top g_S(V)(\Re\mathcal{E})},
\end{equation}
where $\Re\mathcal{E}$ is an unrestricted nonzero real vector. If the real symmetric matrix $g_S(V)$ is positive-definite, Eq.~(\ref{eq:real-rayleigh}) is a generalized Rayleigh quotient and the Courant–Fischer theorem applies~\cite{Lange2010}: it is bounded above and below by the maximum and minimum eigenvalues of the matrix $g_S(V)^{-1}g_S(W)$.

As mentioned above, $g(V)$ is positive-definite. For example, we typically choose $V = \mathbbm{1}_m$, in which case $g_S(V) = \mathbbm{1}_m + F_S F_S^\top$ is necessarily positive-definite.
These results also hold if $g_S(V)$ is negative-definite: by flipping the signs of both $W$ and $V$, one recovers a positive-definite $g_S(-V)$.



\section{Null intensity and skew fraction}\label{app:null-intensity}
For full-rank $g_S(V)$, if $W$ is restricted to a real matrix, the rank of $g_S(V)^{-1}g_S(W)$ satisfies~\cite{Strang2023}
\begin{equation}\label{eq:rank}
\begin{aligned}
&\rank[g_S(V)^{-1}g_S(W)]=\rank[g_S(W)]\\
&=\rank(W+F_SWF_S^\top) \\
&\le \rank(W)+\rank(F_SWF_S^\top)\\
&\le \rank(W)+\min(\rank(F_S),\rank(W))\\
&\le 2\rank(W).
\end{aligned}
\end{equation}
Therefore, if $\rank(W)\le \lfloor (m-1)/2\rfloor$, the matrix $g_S(V)^{-1}g_S(W)$ cannot have full rank, so at least one of its eigenvalues must be zero.

Let us consider a real diagonal matrix $W={\rm diag}(w_1,\cdots,w_m)$ as the choice for Eq.~(\ref{eq:fraction-bounds}). Following the above argument, if the number of nonzero elements in $w_1,\cdots,w_m$ is not greater than $\lfloor (m-1)/2\rfloor$, at least one of the eigenvalues of $g_S(V)^{-1}g_S(W)$ is zero, which allows $\sum_k w_k\eta_k$ to reach zero even if those nonzero weights are all positive or all negative. 
This behavior implies that, given at most $\lfloor (m-1)/2\rfloor$ ports, time-reversal symmetry does not forbid all intensity fractions $\eta_j$ associated with these ports to be zero simultaneously, as exemplified in Fig.~\ref{fig2}(b).

As a complement to the above situation, $g_S(W)$ and hence $g_S(V)^{-1}g_S(W)$ usually have full rank if the real matrix $W$ has $\rank(W)\ge\lfloor (m-1)/2\rfloor+1$, since for generic $S$ one expects $F_S$ to be a full-rank non-sparse matrix. Therefore, given at least $\lfloor (m-1)/2\rfloor+1$ ports, time-reversal symmetry usually forbids intensity fractions associated with these ports to be zero simultaneously at any locations, as exemplified in Figs.~\ref{fig2}(d) and (f).
One can apply similar arguments to $V$: apart from full-rank matrices such as $\mathbbm{1}_m$, setting $V$ as some nearly full-rank matrices may still allow $g_S(V)$ to be positive- or negative-definite. In particular, let $V$ be a diagonal matrix with each element being either zero or one. If at least $\lfloor (m-1)/2\rfloor+1$ of these diagonal elements are equal to one, the matrix $g(V)$ is usually positive-definite. Therefore, in addition to the fraction in Eq.~(\ref{eq:inten-frac}), one can often introduce a skew fraction in which the denominator does not include all the ports and relevant inequalities similar to Eq.~(\ref{eq:fraction-bounds}) can then be formulated.


\section{Effects of unitary transformation}\label{app:unitary-transform}
In this subsection, we discuss the effects of unitary transformations of input and output modes~\cite{Guo2023,Guo2024Absorption,Guo2024Transmission,Guo2024Multiport} on the generalized Rayleigh quotient and its bounds. Under a unitary change $U$ of basis, the $S$~matrix and fields become:
\begin{equation}\label{eq:unitary-transf}
S\rightarrow U^\top SU,~~~~~~\mathcal{E}\rightarrow U^\top \mathcal{E},
\end{equation}
with $U^\dagger U=\mathbbm{1}_m$ and $m$ denoting the number of ports. For example, changing reference planes for the ports amounts to applying a port-wise phase transformation $U=\exp(i\Phi)$, with $\Phi$ being a real diagonal matrix. If modes belong to different ports are mixed, $U$ can be non-diagonal. Time-reversal symmetry requires $S$ to be a coninvolutory matrix, which satisfies $S^*S=\mathbbm{1}_m$~\cite{Haus1984}. The transformed $S$~matrix, namely $U^\top SU$, is still a coninvolutory matrix. The transformed $S$~matrix $U^\top SU$ and transformed fields $U^\top\mathcal{E}$ together still satisfy (\ref{eq:ESE}).

If both $V$ and $W$ are diagonal, as in Eq.~(\ref{eq:fraction-bounds}), both the numerator and denominator of the generalized Rayleigh quotient contain only intensity-like terms that do not depend on the reference planes of the ports. Consequently, the generalized Rayleigh quotient in Eq.~(\ref{eq:Rayleigh-bounds}) does not change under $U=\exp(i\Phi)$ with $\Phi$ being a real diagonal matrix. If $V$, $W$, or both are non-diagonal, cross terms appear, which involve phases differences between fields excited from different ports. Consequently, $U=\exp(i\Phi)$ generally alters the generalized Rayleigh quotient.
Generically, the generalized Rayleigh quotient in Eq.~(\ref{eq:Rayleigh-bounds}) is invariant if $U^\top$ commutes with both $V$ and $W$, namely $U^\top V=VU^\top$ and $U^\top W=WU^\top$. This condition is satisfied, for example, when $U$ performs a global phase transformation, namely $U=\exp(i\Phi)$ with $\Phi\propto\mathbbm{1}_m$.

When the generalized Rayleigh quotient is unchanged under Eq.~(\ref{eq:unitary-transf}) with some choices of $U$, the bounds in Eq.~(\ref{eq:Rayleigh-bounds}) also remain the same. To demonstrate such invariance of bounds, instead of dealing with the bounds directly, one can note that the bounds are always attainable as discussed in Appendix~\ref{app:rayleigh}. Since the bounded quantity does not vary under the transformation and the bounds are attainable, the bounds do not change, either.

As mentioned in Sec.~\ref{sec:trs}, the time-reversal-symmetry bounds in Eq.~(\ref{eq:Rayleigh-bounds}) requires $\mathbbm{1}_m+\Re S$ to be invertible, which may not be satisfied by very unusual examples of $S$,  but can usually be restored by a global phase transformation $U=\exp(i\Phi)$ with some random $\Phi$.

\section{Bounds on a decomposable system}\label{app:sub-systems}

In some situations, typically due to symmetry mismatch or spatial separation, a multi-port system can be divided into two or more subsystems that are decoupled from each other, such that any input to the ports of a subsystem cannot contribute to output in another subsystem, and vice versa, any output from the ports in a subsystem cannot be contributed by input in another subsystem. For example, in 2d systems such as Fig.~\ref{fig2}(a), the $H_z$ and $E_z$ polarizations are decoupled~\cite{Joannopoulos2008}. The $S$~matrix of the entire system is thus block diagonal:
\begin{equation}\label{eq:direct-sum}
S = \bigoplus_{\ell=1}^p S^{(\ell)},
\end{equation}
where $\bigoplus$ denotes the direct sum and $p$ is the number of diagonal blocks of $S$, which is also the number of subsystems. Within each subsystem, Eq.~(\ref{eq:ESE}) holds true as $\mathcal{E}^{(\ell)*}=S^{(\ell)\dagger}\mathcal{E}^{(\ell)}$, where $\mathcal{E}^{(\ell)}$ is the segment of $\mathcal{E}$ associated with this subsystem. If $V$ and $W$ are block-diagonal in the same manner as $S$, one can formulate the inequalities in each subsystem:
\begin{equation}\label{eq:Rayleigh-bounds-sub}
\begin{aligned}
\lambda_{\min}[G_{S^{(\ell)}}(V^{(\ell)},W^{(\ell)})]\le&\frac{\mathcal{E}^{(\ell)\dagger} W^{(\ell)}\mathcal{E}^{(\ell)}}{\mathcal{E}^{(\ell)\dagger} V^{(\ell)}\mathcal{E}^{(\ell)}}\\
\le&\lambda_{\max}[G_{S^{(\ell)}}(V^{(\ell)},W^{(\ell)})].
\end{aligned}
\end{equation}

For the block-diagonal $S$~matrix in Eq.~(\ref{eq:direct-sum}), if $V$ and $W$ are block-diagonal in the same manner, the inequalities~\ref{eq:Rayleigh-bounds-sub} satisfied by each subsystem also imply the inequalities~(\ref{eq:Rayleigh-bounds}) for the total system. In other words, Eq.~(\ref{eq:Rayleigh-bounds}) is not independent of Eq.~(\ref{eq:Rayleigh-bounds-sub}). To see this connection, one can first decompose the generalized Rayleigh quotient for the whole system as
\begin{equation}\label{eq:Rayleigh-totalfromsub}
\frac{\mathcal{E}^\dagger W\mathcal{E}}{\mathcal{E}^\dagger V\mathcal{E}}=\frac{\sum_{\ell=1}^p\mathcal{E}^{(\ell)\dagger} W^{(\ell)}\mathcal{E}^{(\ell)}}{\sum_{\ell=1}^p\mathcal{E}^{(\ell)\dagger} V^{(\ell)}\mathcal{E}^{(\ell)}},
\end{equation}
which is confined by individual subsystems as
\begin{equation}\label{eq:Rayleigh-subboundtotal}
\begin{aligned}
\min_\ell\frac{\mathcal{E}^{(\ell)\dagger} W^{(\ell)}\mathcal{E}^{(\ell)}}{\mathcal{E}^{(\ell)\dagger} V^{(\ell)}\mathcal{E}^{(\ell)}}\le&
\frac{\sum_{\ell=1}^p\mathcal{E}^{(\ell)\dagger} W^{(\ell)}\mathcal{E}^{(\ell)}}{\sum_{\ell=1}^p\mathcal{E}^{(\ell)\dagger} V^{(\ell)}\mathcal{E}^{(\ell)}}\\
\le&\max_\ell\frac{\mathcal{E}^{(\ell)\dagger} W^{(\ell)}\mathcal{E}^{(\ell)}}{\mathcal{E}^{(\ell)\dagger} V^{(\ell)}\mathcal{E}^{(\ell)}}.
\end{aligned}
\end{equation}
Since $S$, $V$, and $W$ have the same block-diagonal structure, so does $G_S(V,W)$, which implies
\begin{equation}\label{eq:bounds-subboundtotal}
\begin{aligned}
&\lambda_{\min}[G_S(V,W)]=\min_\ell\lambda_{\min}[G_{S^{(\ell)}}(V^{(\ell)},W^{(\ell)})],\\
&\lambda_{\max}[G_S(V,W)]=\max_\ell\lambda_{\max}[G_{S^{(\ell)}}(V^{(\ell)},W^{(\ell)})].
\end{aligned}
\end{equation}
On the other hand, Eq.~(\ref{eq:Rayleigh-bounds-sub}) implies
\begin{equation}\label{eq:bounds-subboundsub}
\begin{aligned}
&\min_\ell\lambda_{\min}[G_{S^{(\ell)}}(V^{(\ell)},W^{(\ell)})]\le\min_\ell\frac{\mathcal{E}^{(\ell)\dagger} W^{(\ell)}\mathcal{E}^{(\ell)}}{\mathcal{E}^{(\ell)\dagger} V^{(\ell)}\mathcal{E}^{(\ell)}},\\
&\max_\ell\frac{\mathcal{E}^{(\ell)\dagger} W^{(\ell)}\mathcal{E}^{(\ell)}}{\mathcal{E}^{(\ell)\dagger} V^{(\ell)}\mathcal{E}^{(\ell)}}\le\max_\ell\lambda_{\max}[G_{S^{(\ell)}}(V^{(\ell)},W^{(\ell)})].
\end{aligned}
\end{equation}
From Eqs.~(\ref{eq:Rayleigh-totalfromsub})-(\ref{eq:bounds-subboundsub}), Eq.~(\ref{eq:Rayleigh-bounds}) can be recovered.


\section{Complex linear operation on fields}\label{sec:complex-vector}
If the linear operation $\hat L$ is complex, Eq.~(\ref{eq:ESE}) does not hold and the time-reversal-symmetry bounds usually need to be corrected. In this section, we consider a complex vector field ${\bf J}({\bf x})$ and its overlap with the field ${\bf E}_k({\bf x})$, such that $\mathcal{E}_k=\int d^3{\bf x}\,{\bf J}^*({\bf x})\cdot{\bf E}_k({\bf x})$.

Unless under some special circumstances such as $\Re\bf J\propto\Im\bf J$, the generalized Rayleigh quotient is bounded trivially as
\begin{equation}\label{eq:JE-Rayleigh-bounds}
\lambda_{\min}(V^{-1}W)\le\frac{\mathcal{E}^\dagger W\mathcal{E}}{\mathcal{E}^\dagger V\mathcal{E}}
\le \lambda_{\max}(V^{-1}W),
\end{equation}
where the $S$~matrix and time-reversal symmetry play no role in the bounds, in contrast to the nontrivial inequalities in Eq.~(\ref{eq:fraction-bounds}). To elucidate the attainability of these trivial bounds, we begin with a stronger proposition: an arbitrary complex number $z_k$ can be assigned to each $\mathcal{E}_k$ as $\mathcal{E}_k=z_k$, if the field distributions satisfy
\begin{equation}\label{eq:JE-kth-only}
\begin{aligned}
&\begin{bmatrix}
\mathbbm{1}_m\int d^3{\bf x}\,{\Re}{\bf J}({\bf x})\cdot + F_S^\top\int d^3{\bf x}\,{\Im}{\bf J}({\bf x})\cdot \\ F_S^\top\int d^3{\bf x}\,{\Re}{\bf J}({\bf x})\cdot  -\mathbbm{1}_m\int d^3{\bf x}\,{\Im}{\bf J}({\bf x})\cdot
\end{bmatrix}\Re\mathbbm{E}({\bf x})\\
&=\begin{pmatrix}
\Re Z\\ \Im Z
\end{pmatrix},
\end{aligned}
\end{equation}
where the integrands are understood as the functions given by the dot product; $\mathbbm{E}({\bf x})$ denotes a column vector composed of field distributions, namely $\mathbbm{E}({\bf x})=[{\bf E}_1({\bf x}),\cdots,{\bf E}_m({\bf x})]^\top$; and $Z$ denotes a column vector composed of desired $z_k$ for $k=1,\cdots,m$, namely $Z=(z_1,\cdots,z_m)^\top$, such that $\mathcal{E}=Z$. This equation stems from the relation within $\mathbbm{E}({\bf x})$, namely
\begin{equation}\label{eq:Im-F-Re}
\Im\mathbbm{E}({\bf x})=F_S^\top\Re\mathbbm{E}({\bf x}),
\end{equation}
which resembles Eq.~(\ref{eq:ESE-split}). Splitting ${\bf J}({\bf x})$ and $\mathbbm{E}({\bf x})$ into real and imaginary parts and using Eq.~(\ref{eq:Im-F-Re}), one can obtain
\begin{equation}
\begin{aligned}
&\int d^3{\bf x}\,{\bf J}({\bf x})\cdot\mathbbm{E}({\bf x})\\
&=\int d^3{\bf x}\,\Big\{{\Re}{\bf J}({\bf x})\cdot\Re\mathbbm{E}({\bf x})+{\Im}{\bf J}({\bf x})\cdot\Im\mathbbm{E}({\bf x})\\
&~~~~~~~~~~+i\big[{\Re}{\bf J}({\bf x})\cdot\Im\mathbbm{E}({\bf x})-{\Im}{\bf J}({\bf x})\cdot\Re\mathbbm{E}({\bf x})\big]\Big\}\\
&=\int d^3{\bf x}\,\Big\{{\Re}{\bf J}({\bf x})\cdot\Re\mathbbm{E}({\bf x})+{\Im}{\bf J}({\bf x})\cdot F_S^\top\Re\mathbbm{E}({\bf x})\\
&~~~~~~~~~~+i\big[{\Re}{\bf J}({\bf x})\cdot F_S^\top\Re\mathbbm{E}({\bf x})-{\Im}{\bf J}({\bf x})\cdot\Re\mathbbm{E}({\bf x})\big]\Big\}\\
&=\left[\begin{pmatrix}
\mathbbm{1}&F_S^\top
\end{pmatrix}+i\begin{pmatrix}
F_S^\top&-\mathbbm{1}
\end{pmatrix}\right]\int d^3{\bf x}\,\begin{bmatrix}
{\Re}{\bf J}({\bf x})\cdot\Re\mathbbm{E}({\bf x})\\
{\Im}{\bf J}({\bf x})\cdot\Re\mathbbm{E}({\bf x})
\end{bmatrix}.
\end{aligned}
\end{equation}
Therefore, $\int d^3{\bf x}\,{\bf J}({\bf x})\cdot\mathbbm{E}({\bf x})=Z$ is equivalent to
\begin{equation}\label{eq:JE-kth-only-2}
\begin{aligned}
&\begin{pmatrix}
\mathbbm{1}&F_S^\top
\end{pmatrix}\int d^3{\bf x}\,\begin{bmatrix}
{\Re}{\bf J}({\bf x})\cdot\Re\mathcal{E}({\bf x})\\
{\Im}{\bf J}({\bf x})\cdot\Re\mathcal{E}({\bf x})
\end{bmatrix}=\Re Z,\\
&\begin{pmatrix}
F_S^\top&-\mathbbm{1}
\end{pmatrix}\int d^3{\bf x}\,\begin{bmatrix}
{\Re}{\bf J}({\bf x})\cdot\Re\mathcal{E}({\bf x})\\
{\Im}{\bf J}({\bf x})\cdot\Re\mathcal{E}({\bf x})
\end{bmatrix}=\Im Z,
\end{aligned}
\end{equation}
which can be written compactly as Eq.~(\ref{eq:JE-kth-only}). 

As an approximation of the overlap integral, one can consider a discretized version as $\mathcal{E}_k=\sum_{\ell=1}^n{\bf J}^*({\bf x}_\ell)\cdot{\bf E}_k({\bf x}_\ell)$, with the volume factor in the integral being discarded and $n$ representing the number of locations. Analogues to Eq.~(\ref{eq:JE-kth-only}), $\mathcal{E}=Z$ can be attained if the field distributions satisfy
\begin{equation}\label{eq:JE-kth-only-discrete}
\begin{aligned}
\begin{pmatrix}
\mathbbm{1}_m\otimes\Re\mathcal{J}^\top+F_S^\top\otimes\Im\mathcal{J}^\top \\
F_S^\top\otimes\Re\mathcal{J}^\top-\mathbbm{1}_m\otimes\Im\mathcal{J}^\top
\end{pmatrix}\Re\mathbbm{E}=\begin{pmatrix}
\Re Z\\ \Im Z
\end{pmatrix},
\end{aligned}
\end{equation}
where we define $\mathcal{J}$ as a column vector that contains each component of $\bf J$ at each location, and $\mathbbm{E}$ as a column vector that contains each component of ${\bf E}_j$ at each location for $j=1,\cdots,m$, such that
\begin{equation}\label{eq:JE-integral}
\mathcal{J}^\dagger\mathbbm{E}=\sum_{\ell=1}^n\begin{bmatrix}
{\bf J}^*({\bf x}_\ell)\cdot{\bf E}_1({\bf x}_\ell)\\
\vdots\\
{\bf J}^*({\bf x}_\ell)\cdot{\bf E}_m({\bf x}_\ell)
\end{bmatrix}.
\end{equation}
One can also expand ${\bf J}^*({\bf x})$ and each ${\bf E}_k({\bf x})$ in terms of functions that form an orthonormal basis. With $\mathcal{J}$ and $\mathbbm{E}$ understood as column vectors composed of the expansion coefficients, Eq.~(\ref{eq:JE-kth-only-discrete}) can be regarded as an approximation or alternative to Eq.~(\ref{eq:JE-kth-only}).

If $\mathcal{J}$ contains at least three nonzero elements while this column vector and the $S$~matrix are not too special, such that the coefficient matrix in the parentheses on the left-hand side of Eq.~(\ref{eq:JE-kth-only-discrete}) has full row rank but does not have full column rank, this linear equation is underdetermined, which has infinitely many solutions for $\Re\mathbbm{E}$, from which one can compute the corresponding imaginary part $\Im\mathbbm{E}$ using Eq.~(\ref{eq:Im-F-Re}). If $\mathcal{J}$ contains only two elements while this vector and the $S$~matrix are not too special, such that the coefficient matrix has both full row rank and column rank, this linear equation has a single solution for $\Re\mathbbm{E}$. In special cases such as $\Re\bf J\propto\Im\bf J$, this linear equation generally has no solution unless the $S$~matrix and $Z$ are also special enough.

As discussed above, in a typical situation, assigning an arbitrary set of values $Z=(z_1,\cdots,z_m)^\top$ to $\mathcal{E}$ are permitted by time-reversal symmetry. If $\mathcal{E}$ is set as the eigenvector of $V^{-1}W$ corresponding to the minimum or maximum eigenvalue, the trivial lower or upper bound in Eq.~(\ref{eq:JE-Rayleigh-bounds}) is attained. Although allowed by time-reversal symmetry, these trivial bounds may not be reached given a specific structure and a specific complex vector field $\bf J$, even if $\Re\bf J\propto\Im\bf J$ is avoided, because the field distributions $\mathbbm{E}$, which observe the Maxwell equations, cannot be chosen arbitrarily.

Although time-reversal symmetry alone generally cannot impose nontrivial bounds on the generalized Rayleigh quotient for $\int d^3{\bf x}\,{\bf J}^*({\bf x})\cdot{\bf E}_k({\bf x})$ in the fashion as Eq.~(\ref{eq:Rayleigh-bounds}), time-reversal symmetry may still impose nontrivial bounds on some related quantities. 
For example, consider a ratio defined as
\begin{equation}
\rho_k = \frac{\left|\int d^3{\bf x}\,{\bf J}^*({\bf x})\cdot{\bf E}_k({\bf x})\right|^2}{\sum_{j=1}^m \int d^3{\bf x}\,{\bf E}_j({\bf x})\cdot{\bf D}_j({\bf x})},
\end{equation}
where the domains of the integrals are the same, and the media are assumed to have real symmetric positive-definite permittivity tensors.
One can introduce a discretized version as
\begin{equation}
\rho_k = \frac{\left|\sum_{\ell=1}^n{\bf J}^*({\bf x}_\ell)\cdot{\bf E}_k({\bf x}_\ell)\right|^2}{\sum_{j=1}^m\sum_{\ell=1}^n{\bf E}_j({\bf x}_\ell)\cdot{\bf D}_j({\bf x}_\ell)},
\end{equation}
where the subscript $\ell$ runs over all locations and directions. This fraction can be bounded in the same manner as Eq.~(\ref{eq:fraction-bounds}), with $W={\rm diag}(w_1,\cdots,w_m)$ and $G_S$ being replaced by
\begin{equation}\label{eq:JE-bound-matrix}
(\mathcal{J}_+W\mathcal{J}_+^\top+\mathcal{J}_-W\mathcal{J}_-^\top)[(F_SF_S^\top+\mathbbm{1}_m)\otimes\Xi]^{-1},
\end{equation}
where $\Xi$ is a diagonal or block-diagonal matrix containing the relative permittivity tensor for each location and direction, and
we also introduce
\begin{equation}
\begin{aligned}
&\mathcal{J}_+=\mathbbm{1}_m\otimes\Re\mathcal{J}+F_S\otimes\Im\mathcal{J},\\
&\mathcal{J}_-=F_S\otimes\Re\mathcal{J}-\mathbbm{1}_m\otimes\Im\mathcal{J},
\end{aligned}
\end{equation}
where $\mathcal{J}$ represents the column vector for the components of $\bf J$ at all relevant locations, with the same meaning as that in Eqs.~(\ref{eq:JE-kth-only-discrete}) and (\ref{eq:JE-integral}).

If the linear operation $\hat L$ is simply scalar multiplication by a complex number, Eq.~(\ref{eq:ESE}) only needs to slightly tweaked and all the time-reversal-symmetry inequalities in the main text still hold true.

~\\
\section{Bounds on emitted waves}\label{app:emit}
In this appendix, we discuss in more detail the emission from an internal point dipole source and the time-reversal-symmetry bounds on the amplitudes, presented in Sec.~\ref{subsec:emit}.
Let ${\bf E}$, ${\bf D}$, ${\bf H}$, and ${\bf B}$ denote the complex amplitudes of electric, displacement, magnetizing, and magnetic fields, respectively. The constitutive relations are
\begin{equation}
\begin{aligned}
&{\bf D}({\bf r},\omega)={\epsilon}({\bf r},\omega){\bf E}({\bf r},\omega)+\xi({\bf r},\omega){\bf H}({\bf r},\omega),\\
&{\bf B}({\bf r},\omega)={\zeta}({\bf r},\omega){\bf E}({\bf r},\omega)+\mu({\bf r},\omega){\bf H}({\bf r},\omega),
\end{aligned}
\end{equation}
where $\epsilon$ and $\mu$ are permittivity and permeability tensors, respectively; $\xi$ and $\zeta$ are tensors describing magnetoelectric couplings~\cite{Mackay2019}. The two Maxwell equations with curl are
\begin{equation}\label{eq:Maxwell}
\begin{aligned}
&\nabla\times{\bf H}+i\omega(\epsilon{\bf E}+\xi{\bf H}) = {\bf J}_{\rm e},\\
&\nabla\times{\bf E}-i\omega(\zeta{\bf E}+\mu{\bf H}) = -{\bf J}_{\rm m},
\end{aligned}
\end{equation}
where ${\bf J}_{\rm e}$ and ${\bf J}_{\rm m}$ are electric and magnetic current densities, respectively.

On the other hand, in a complementary system with~\cite{JinAuKong1972,JinAuKong2008}
\begin{equation}\label{eq:tensors-T}
\Tilde{\epsilon}=\epsilon^\top,~~~~\Tilde{\mu}=\mu^\top,~~~~\Tilde{\xi}=-\zeta^\top,~~~~\Tilde{\zeta}=-\xi^\top,
\end{equation}
fields and currents satisfy
\begin{equation}\label{eq:Maxwell-T}
\begin{aligned}
&\nabla\times\Tilde{\bf H}+i\omega(\Tilde{\epsilon}\Tilde{\bf E}+\Tilde{\xi}\Tilde{\bf H}) = \Tilde{\bf J}_{\rm e},\\
&\nabla\times\Tilde{\bf E}-i\omega(\Tilde{\zeta}\Tilde{\bf E}+\Tilde{\mu}\Tilde{\bf H}) = -\Tilde{\bf J}_{\rm m}.
\end{aligned}
\end{equation}
From Eqs.~(\ref{eq:Maxwell}), (\ref{eq:tensors-T}), and (\ref{eq:Maxwell-T}), after some algebra, one can obtain
\begin{equation}\label{eq:general-recipr}
\begin{aligned}
&\varoiint_{\partial\mathcal{V}}({\bf E}\times\Tilde{\bf H}-\Tilde{\bf E}\times{\bf H})\cdot d{\bf A}=\\
&\iiint_{\mathcal{V}}
(\Tilde{\bf E}\cdot{\bf J}_{\rm e}+{\bf H}\cdot\Tilde{\bf J}_{\rm m}-{\bf E}\cdot\Tilde{\bf J}_{\rm e}-\Tilde{\bf H}\cdot{\bf J}_{\rm m})\,d\mathcal{V},
\end{aligned}
\end{equation}
where $\mathcal{V}$ and $\partial\mathcal{V}$ denote a region and its surface, respectively, and $d{\bf A}$ is the surface-element vector parallel to the normal pointing outward. If the only impressed current in the region $\mathcal{V}$ is $\Tilde{\bf J}_{\rm e}$, the above relation becomes
\begin{equation}\label{eq:single-recipr}
\varoiint_{\partial\mathcal{V}}({\bf E}\times\Tilde{\bf H}-\Tilde{\bf E}\times{\bf H})\cdot d{\bf A}=-\iiint_{\mathcal{V}}{\bf E}\cdot\Tilde{\bf J}_{\rm e}\,d\mathcal{V},
\end{equation}
where $\Tilde{\bf E}$ and $\Tilde{\bf H}$ are excited by $\Tilde{\bf J}_{\rm e}$.

Let the boundary $\partial\mathcal{V}$ intersect the input and output channels and be distant from the scattering media. If the scattering device is a one- or two-dimensional periodic structure, $\partial\mathcal{V}$ can be chosen as a pair of planes on either side of the periodic structure.
As a typical scenario, we assume that the media far from the scattering region to have real symmetric $\epsilon$, $\mu$, and $\xi=\zeta=0$, and that the geometry of each channel has translational invariance in the direction of mode propagation.
In the original structure (without tildes in labels of fields), when the input only comes from the $k$-th port with a unit amplitude, the fields at $\partial\mathcal{V}$ can be written as
\begin{equation}
{\bf E}={\bf E}_k^{\rm in}+\sum_{j=1}^m c_j{\bf E}_j^{\rm out},~~~~{\bf H}={\bf H}_k^{\rm in}+\sum_{j=1}^m c_j{\bf H}_j^{\rm out},
\end{equation}
where each $c_j$ represents a complex coefficient. On the other hand, the input and output modes are related as
\begin{equation}
{\bf E}_k^{\rm in} = {\bf E}_k^{{\rm out}*},~~~~~~{\bf H}_k^{\rm in} = -{\bf H}_k^{{\rm out}*}.
\end{equation}
Therefore, the left-hand side of Eq.~(\ref{eq:single-recipr}) becomes
\begin{equation}\label{eq:mode-overlap}
\begin{aligned}
&\varoiint_{\partial\mathcal{V}}({\bf E}\times\Tilde{\bf H}-\Tilde{\bf E}\times{\bf H})\cdot d{\bf A}\\
&=\varoiint_{\partial\mathcal{V}}({\bf E}_k^{{\rm out}*}\times\Tilde{\bf H}+\Tilde{\bf E}\times{\bf H}_k^{{\rm out}*})
\cdot d{\bf A}\\
&~~~~+\sum_{j=1}^m c_j\varoiint_{\partial\mathcal{V}} ({\bf E}_j^{{\rm in}*}\times\Tilde{\bf H}+\Tilde{\bf E}\times{\bf H}_j^{{\rm in}*})\cdot d{\bf A}.
\end{aligned}
\end{equation}
Since $\Tilde{\bf E}$ and $\Tilde{\bf H}$ are the fields of the outgoing wave in the far-field region, the second term on the right-hand side of Eq.~(\ref{eq:mode-overlap}) vanishes, while the remaining first term is proportional to the $k$-th output mode coefficient, which is denoted by $A_k$ hereafter.

In particular, when $\Tilde{\bf J}_{\rm e}$ is a point dipole source located at $\bf r$ oriented along a unit vector $\hat{n}$, i.e., $\Tilde{\bf J}_{\rm e}\propto{\hat{n}}\,\delta({\bf x}-{\bf r})$, the field component ${\bf E}({\bf r})\cdot{\hat{n}}$ is singled out on the right-hand side of Eq.~(\ref{eq:single-recipr}). (If the scattering device is a one- or two-dimensional periodic structure, $\Tilde{\bf J}_{\rm e}$ is understood as a periodic array of point dipole sources.) Due to the assumption that the input only comes from the $k$-th channel with a unit amplitude, one can replace ${\bf E}({\bf r})\cdot{\hat{n}}$ here with ${\bf E}_k({\bf r})\cdot{\hat{n}}$ and obtain
\begin{equation}
A_k\propto{\bf E}_k({\bf r})\cdot{\hat{n}}.
\end{equation}
Similarly, the output power $P_k$ obeys
\begin{equation}
P_k\propto|A_k|^2\propto|{\bf E}_k({\bf r})\cdot{\hat{n}}|^2.
\end{equation}
If the orientation of the point dipole source is completely random, the average emitted power carried by that mode is
\begin{equation}
\overline{P_k}\propto\overline{|{\bf E}_k({\bf r})\cdot{\hat{n}}|^2}=\frac{|{\bf E}_k({\bf r})|^2}{3},
\end{equation}
where the bar denotes average over $\hat{n}$.
Consequently, $A_k$, $P_k$, and $\overline{P_k}$ in the complementary system satisfy the same time-reversal-symmetry bounds as the fields in the original system.  Conversely, the time-reversal-symmetry bounds on emitted-wave amplitudes from a point dipole source in the original system should coincide with the bounds on fields in the complementary system. The $S$~matrix of the complementary system is $S^\top$ if the $S$~matrix of the original system is $S$. Therefore, time-reversal-symmetry bounds on emitted-wave amplitudes from a point dipole source can be obtained immediately from the corresponding bounds for fields via replacing $S$ with $S^\top$. The $S$~matrix is symmetric if the media are reciprocal, requiring
\begin{equation}
\epsilon=\epsilon^\top,~~~~\mu=\mu^\top,~~~~\xi=-\zeta^\top,~~~~\zeta=-\xi^\top.
\end{equation}
Therefore, the time-reversal-symmetry bounds on the fields induced by input coincide with those on emitted-wave amplitudes from a point dipole source.

~\\
\section{Bounds on resonant coupling and decay coefficients}\label{app:TCMT}

Finally, in this appendix, we prove the bounds from Sec.~\ref{sec:tcmt} for coupled-resonant systems described by TCMT. When excited at the frequency $\omega$, the resonant amplitude is
\begin{equation}\label{eq:res-amplitude}
a = \frac{\bm{\kappa}^\top{\bf{s}}_+}{\gamma+i(\omega-\omega_0)}.
\end{equation}
Given the input wave only from the $k$-th port with a unit amplitude, the resonant amplitude is
\begin{equation}
a_k = \frac{\kappa_{k}}{\gamma+i(\omega-\omega_0)},
\end{equation}
which is proportional to the magnitude of ${\bf{E}}_k$ at a given location in the resonator and a given frequency, and further implies
\begin{equation}
\frac{\mathcal{E}^\dagger W\mathcal{E}}{\mathcal{E}^\dagger V\mathcal{E}}=\frac{\bm{\kappa}^\dagger W\bm{\kappa}}{\bm{\kappa}^\dagger V\bm{\kappa}},
\end{equation}
where each component of $\mathcal{E}$ can be considered as ${\bf{E}}_k$.
Therefore, similar to Appendix~\ref{app:rayleigh}, the generalized Rayleigh quotient with ${\bm\kappa}$ also satisfies the bounds in Eq.~(\ref{eq:Rayleigh-bounds}), in which $S$ is the overall $S$~matrix:
\begin{equation}\label{eq:S-tcmt}
S = C+\frac{{\bf{d}}\bm{\kappa}^\top}{\gamma+i(\omega-\omega_0)}.
\end{equation}
For $|\omega-\omega_0|\gg\gamma$, the term for resonance in Eq.~(\ref{eq:S-tcmt}), namely the second term on the right-hand side, vanishes and $S$ can be approximated by $C$. Let us assume that such $\omega$ is still in the frequency range over which $C$, $\bf{d}$, $\bm{\kappa}$, and $\gamma$ do not change and TCMT is still valid. The bounds for $\bm{\kappa}$ can then be expressed in terms of $C$ instead of $S$ and the first lines of Eqs.~(\ref{eq:Rayleigh-bounds-tcmt}) and (\ref{eq:rate-bounds-tcmt}) can be obtained.

The analysis of resonant decay is similar. Based on Eqs.~(\ref{eq:Rayleigh-bounds-emit}) and (\ref{eq:fraction-bounds-emit}), the second lines of Eqs.~(\ref{eq:Rayleigh-bounds-tcmt}) and (\ref{eq:rate-bounds-tcmt}) can be obtained. We obtain Eqs.~(\ref{eq:Rayleigh-bounds-emit}) and (\ref{eq:fraction-bounds-emit}) under the assumption of sustained excitation, which is approximately fulfilled if the time scale of resonant decay $1/\gamma$ is much longer than the period of waves $1/\omega_0$, namely $\omega_0\gg\gamma$, which is a typical premise of TCMT. 

The same results can alternatively be derived from the time-reversal-symmetry requirements in TCMT~\cite{Zhao2019}:
\begin{equation}\label{eq:cdck-tcmt}
C^\top{\bm{\kappa}}^*=-\bm{\kappa},~~~~~~C{\bf{d}}^*=-{\bf{d}}.
\end{equation}
For multiple resonant modes coupled with each other and with multiple ports~\cite{Suh2004}, Eq.~(\ref{eq:cdck-tcmt}) holds for each resonant mode, and hence the bounds are applicable to each resonant mode coupled with multiple ports.



\providecommand{\noopsort}[1]{}\providecommand{\singleletter}[1]{#1}%

\end{document}